\documentclass[pdflatex,sn-mathphys-num]{sn-jnl}

\usepackage{graphicx}%
\usepackage{multirow}%
\usepackage{amsmath,amssymb,amsfonts}%
\usepackage{amsthm}%
\usepackage{mathrsfs}%
\usepackage[title]{appendix}%
\usepackage{xcolor}%
\usepackage{textcomp}%
\usepackage{manyfoot}%
\usepackage{booktabs}%
\usepackage{algorithm}%
\usepackage{algorithmicx}%
\usepackage{algpseudocode}%
\usepackage{listings}%
\usepackage{longtable}%
\usepackage{subcaption}

\theoremstyle{thmstyleone}%
\theoremstyle{thmstyletwo}%

\theoremstyle{thmstylethree}%

\newcommand{\msun}{\mbox{M$_{\odot}$}}
\newcommand{\rsun}{\mbox{R$_{\odot}$}}
\newcommand{\lsun}{\mbox{L$_{\odot}$}}

\newcommand{\teff}{\textit{T$_{\rm eff}$}}

\newcommand{\Mdot}{$\dot{M}$}
\usepackage{lineno}

\begin{document}

\title[Article Title]{The dramatic transition of the extreme Red Supergiant WOH G64 to a Yellow Hypergiant}


\author*[1,2]{\fnm{Gonzalo} \sur{Mu\~noz-Sanchez}}\email{gonzalomu16@gmail.com}

\author[1,2]{\fnm{Maria} \sur{Kalitsounaki}}

\author[1,2]{\fnm{Stephan} \sur{de Wit}}

\author[1,2]{\fnm{Konstantinos} \sur{Antoniadis}}

\author[1,2]{\fnm{Alceste Zoe} \sur{Bonanos}}

\author[2,3]{\fnm{Emmanouil} \sur{Zapartas}}

\author[4,5]{\fnm{Konstantina} \sur{Boutsia}}

\author[1,2]{\fnm{Evangelia} \sur{Christodoulou}}

\author[1,3]{\fnm{Grigoris} \sur{Maravelias}}

\author[6]{\fnm{Igor} \sur{Soszy\'nski}}

\author[6]{\fnm{Andrzej} \sur{Udalski}}

\affil[1]{\orgdiv{IAASARS}, \orgname{National Observatory of Athens}, \orgaddress{\street{I. Metaxa \& Vas. Pavlou St.}, \city{Penteli}, \postcode{15236}, \state{Athens}, \country{Greece}}}

\affil[2]{\orgdiv{Department of Physics}, \orgname{National and Kapodistrian University of Athens}, \orgaddress{\street{Panepistimiopolis}, \city{Zografos}, \postcode{15784}, \state{Athens}, \country{Greece}}}

\affil[3]{\orgdiv{Institute of Astrophysics}, \orgname{FORTH}, \orgaddress{\city{Heraklion}, \postcode{71110}, \state{Crete}, \country{Greece}}}

\affil[4]{\orgdiv{Cerro Tololo Inter-American Observatory/NSF}, \orgname{NOIRLab}, \orgaddress{\street{Casilla 603}, \state{La Serena}, \country{Chile}}}

\affil[5]{\orgdiv{Las Campanas Observatory}, \orgname{Carnegie Observatories}, \orgaddress{\street{Colina El Pino}, \city{Casilla 601}, \state{La Serena}, \country{Chile}}}

\affil[6]{\orgdiv{Astronomical Observatory}, \orgname{University of Warsaw}, \orgaddress{\street{Al. Ujazdowskie 4}, \postcode{00-478}, \state{Warszawa}, \country{Poland}}}

\abstract{Red Supergiants (RSGs) are cool, evolved massive stars in their final evolutionary stage before exploding as a supernova. However, the evolution and fate of the most luminous RSGs remain uncertain. Observational evidence for luminous warm, post-RSG objects and the apparent lack of luminous RSGs as supernova progenitors suggest a blueward evolution. Since the 1980s, WOH~G64 has been considered the most extreme RSG in the Large Magellanic Cloud, given its large obscuration, outstanding size, luminosity, and mass-loss rate. Here we report a sudden, yet smooth change in its apparent nature. Time-series photometry and subsequent spectroscopy revealed the most extreme transition ever seen in the optical spectral features of a RSG. We discovered that WOH~G64 is a rare, massive symbiotic binary system where the RSG transitioned to a Yellow Hypergiant. The dramatic transition can be explained either by the partial ejection of the pseudo-atmosphere during a common-envelope phase, or the return to a quiescent state after an outstanding eruption exceeding 30 years. WOH~G64 offers a unique opportunity to witness stellar evolution in real-time and assess the role of binarity on the final phases of massive stars and their resulting supernovae.}




\maketitle


Red Supergiants (RSGs) are cool, evolved massive stars with initial masses 8--30~\msun~representing the final evolutionary stage before ending their lives in a Type~II supernova (SN) explosion \citep{Meynet2003, Ekstrom2012, Levesque2017, VanDyk2025}. However, the apparent lack of luminous RSGs detected as SN progenitors has sparked an ongoing debate over the fate of these stars. While several studies support the existence of the ``RSG problem"  \cite{Li2006, Kochanek2008, Smartt2009II, Smartt2015, Kochanek2020, Fang2025}, others argue that it is not statistically significant and may stem from circumstellar extinction and assumptions regarding bolometric corrections when inferring SN progenitor properties \cite{Davies2020_luminosity, Davies2020_RSGproblem, Beasor2024, Strotjohann2024, Healy2024}. Alternative explanations include the failed SN scenario by the direct collapse into a black hole \cite{De2024}, or post-RSG evolution \citep[i.e., Yellow Hypergiant (YHG) phase;][]{deJager1998, Gordon2019, Humphreys2022, Jones2025} in which the star loses its outer envelope. The scarcity of very luminous RSGs, the presence of large amounts of circumstellar material (CSM), and their inaccurate distances \citep{Humphreys2019} hinder accurate measurements of their stellar properties, complicating the assessment of their evolutionary stage and ultimate fate.

Mass loss is one of the key parameters shaping the evolution and fate of luminous RSGs. Strong, steady winds could strip their stellar envelope, triggering a post-RSG evolution phase \citep{Meynet2015, Zapartas2024} and potentially explaining the formation of stripped-envelope SN progenitors \citep{Smith2014}. However, recent empirical studies \citep[e.g.,][]{Beasor2020, Antoniadis2024, Decin2024, Antoniadis2025} derived weak winds, which are insufficient to produce post-RSG objects. The detection of vast episodic mass-loss events in luminous RSGs (e.g., NML Cyg and VY CMa) \citep{Richards1996, Decin2006, Singh2023} provides an alternative mechanism to the steady winds. However, the significance of episodic mass-loss drastically varies among RSGs \citep{Humphreys2022}. Still, the physical mechanism driving these events and their impact on stellar evolution remains unknown. Eruptions and large mass ejections due to instabilities \citep{Glatzel1993} are expected at luminosities close to the Humphreys-Davidson limit, $\log(L/\lsun)=5.5-5.8$ \citep{Humphreys1979, Davies2020_luminosity, McDonald2022,Schootemeijer2025}. Alternatively, interactions with a companion can trigger episodic mass-loss events \citep{Landri2024, MacLeod2024}. Binarity adds further complexity to the fate of RSG as it can drive stellar evolution by entering a common-envelope phase (CE), stripping their envelope, or leading to a merger \citep{Sana2012, Ivanova2013}.

WOH~G64 (IRAS 04553-6825) \cite{Westerlund1981} stands out as the most extreme RSG in the Large Magellanic Cloud (LMC), on the brink of the luminosity limit, $\log(L/\lsun)=5.45\pm 0.05$ \cite{Ohnaka2008}, being one of the coolest (M5-7e) \cite{Elias1986, Levesque2009}, largest ($R=1540$~\rsun) \cite{Levesque2009} and the record-holder in terms of mass-loss rate (\Mdot~$>10^{-4}$ \msun~yr$^{-1}$) \cite{Beasor2022, Antoniadis2024}. Interferometric observations in the near and mid-IR revealed elongated emission, which was attributed to a dusty, disk-like structure with a gaseous envelope of 3-9~\msun~\citep{Ohnaka2008, Ohnaka2024}. It is also an OH/IR star with strong multi-component maser emission, possibly due to an external disk, two expanding dust shells, or bipolar outflows \citep{vanLoon2001, Marshall2004}. In this work, we present new observations of WOH~G64, combined with archival photometric and spectroscopic data, to analyze its current evolutionary stage.

\section*{Results}

\begin{figure}[h]
   	\resizebox{1.0\hsize}{!}{\includegraphics{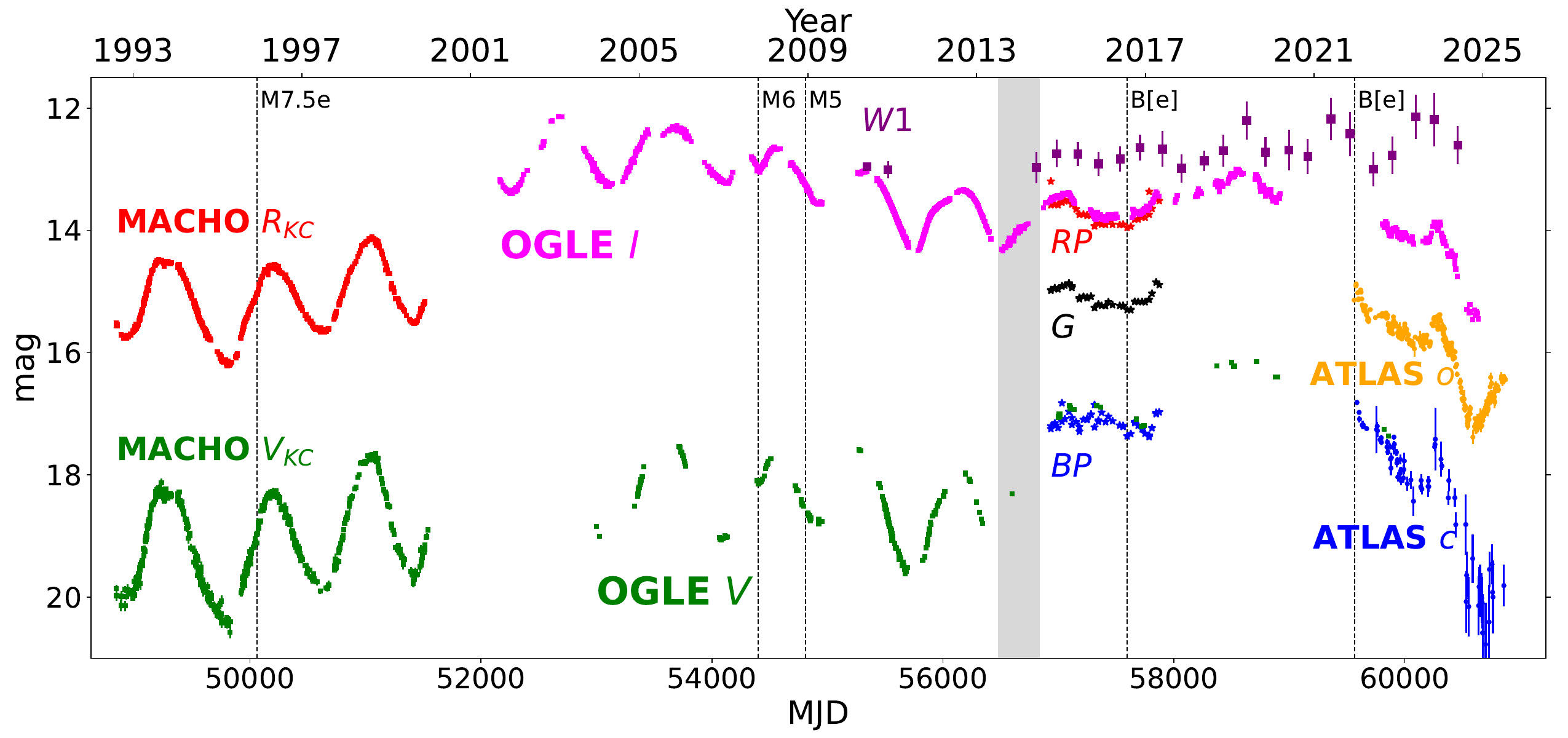}}
        \caption{\textbf{Light curves of WOH~G64}. Each dataset is labeled with the corresponding survey and filter. NEOWISE-R W1 data is shown with a +7.4~mag offset. The gray patch between 2013.5 and 2014.5 constrains the epoch of the transition. The vertical lines indicate the epochs of spectroscopy and the corresponding optical spectroscopic classifications reported for WOH~G64 (Table~\ref{tab:spectral_class}).}
   \label{fig:fig_lightcurve}
\end{figure}

\begin{figure}[H]
\centering
\resizebox{1.\hsize}{!}{\includegraphics{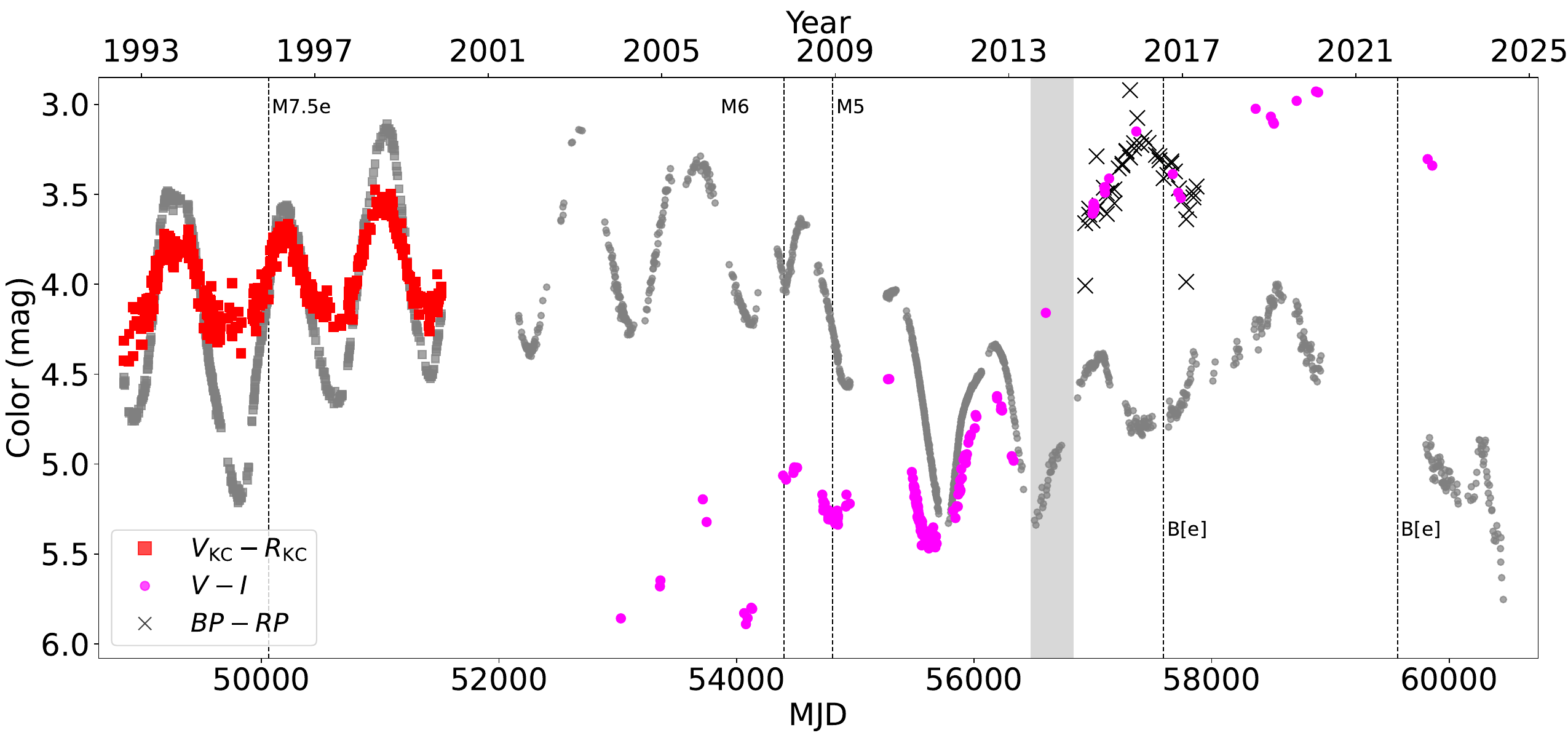}}
        \caption{\textbf{Evolution of the optical colors of WOH~G64}. MACHO $V_{\rm KC}-R_{\rm KC}$ is shown in red squares, OGLE $V-I$ in magenta circles, and \textit{Gaia} $BP-RP$ in black crosses. $R_{\rm KC}$ and $I$ are represented in gray squares and circles, respectively, with an offset for comparison. Vertical dashed lines indicate the optical spectral classifications. The estimated time of the transition between 2013.5 and 2014.5 is highlighted with the shaded area.}
   \label{fig:fig_color_change}
\end{figure}

The optical light curve of WOH~G64, spanning from 1992 to the present, clearly shows two distinct phases, separated by a transition in 2014 (Fig.~\ref{fig:fig_lightcurve}). Before this transition, WOH~G64 was classified as a Mira variable \cite{Soszynski2009} due to its semi-regular periodicity of approximately 850~days, with an amplitude of $\Delta V \thickapprox 2$~mag and $\Delta I \thickapprox 1.5$~mag. However, the sinusoidal pattern observed in the 1990s does not exhibit the typical loop seen in pulsating stars in a color–magnitude diagram (Extended Data Fig.~\ref{fig:color_MARCS}) \cite{Boyd2021}. Instead, MACHO $V_{\rm KC}$ and $R_{\rm KC}$ show a linear trend in the color–magnitude diagram, which can be attributed to semi-periodic extinction ($A_V$) variations of several magnitudes (see Methods). The lack of a correlation between the spectral type and brightness of WOH~G64 during its RSG phase (Fig.~\ref{fig:fig_lightcurve}) agrees with $A_V$ primarily driving the photometric variability \citep{Kravchenko2019}. The strong dimming event in 2011, before the transition, exhibits a loop in the color-magnitude diagram, suggesting variations in the intrinsic properties of the system. After this dimming event, the star became significantly bluer between mid-2013 and mid-2014, with a color change of $\Delta(V-I) \thickapprox 1.8$~mag (Fig.~\ref{fig:fig_color_change}). The change was caused exclusively by a sudden brightening in the $V$-band within a single year, while the $I$-band remained approximately constant. An increase of the $T_{\rm eff}$ by more than 1,000~K explains the photometric variations in the color–magnitude diagram during the transition (Extended Data Fig.~\ref{fig:color_MARCS}). Since 2014, WOH~G64 has shown irregular variability, including a fading of 2~mag over 10 months during 2025, reaching its faintest observed $I$-band state to date. Mid-infrared (IR) brightness before and after the transition does not show significant long-term changes, except for three brightenings with $\Delta W1 \thickapprox 0.5$~mag and a potential periodicity of $\sim$~750 days. Two of these have corresponding brightenings in the optical bands. The large uncertainties in the W1 data limit further analysis.


\begin{figure}[h]
   	\resizebox{1.0\hsize}{!}{\includegraphics{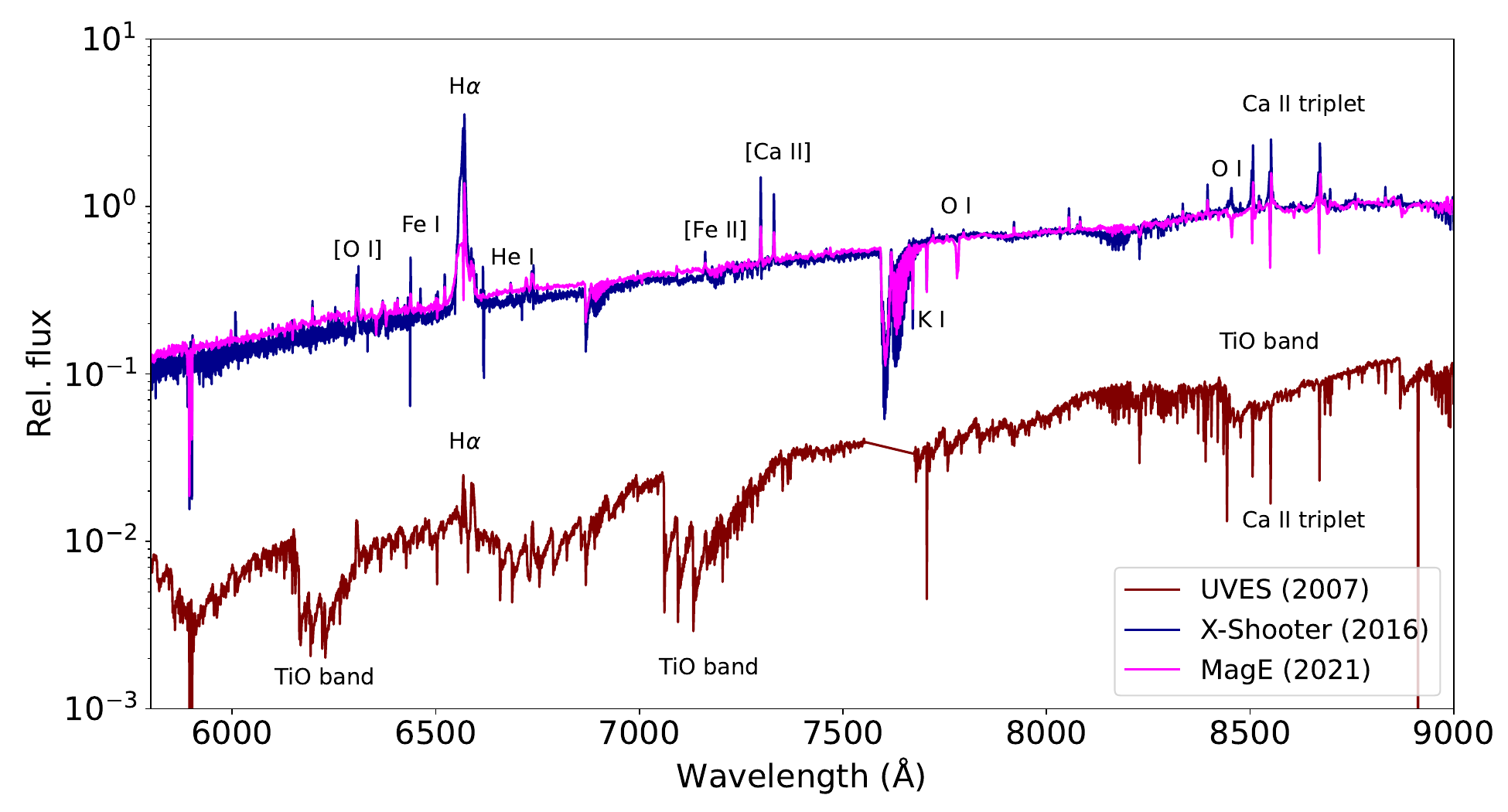}}
        \caption{\textbf{Optical spectra of WOH~G64 illustrating its dramatic transition from a late-M star to a B[e] star.} The main spectral features are indicated for each epoch. The spectra were scaled to 1 between 8700-8800~\r{A}. The UVES spectrum is shown with an offset and downgraded to a spectral resolution of 4000.}
   \label{fig:fig_spec_comparison}
\end{figure}

\begin{figure}[H]
    \centering
    \resizebox{1.0\hsize}{!}{\includegraphics{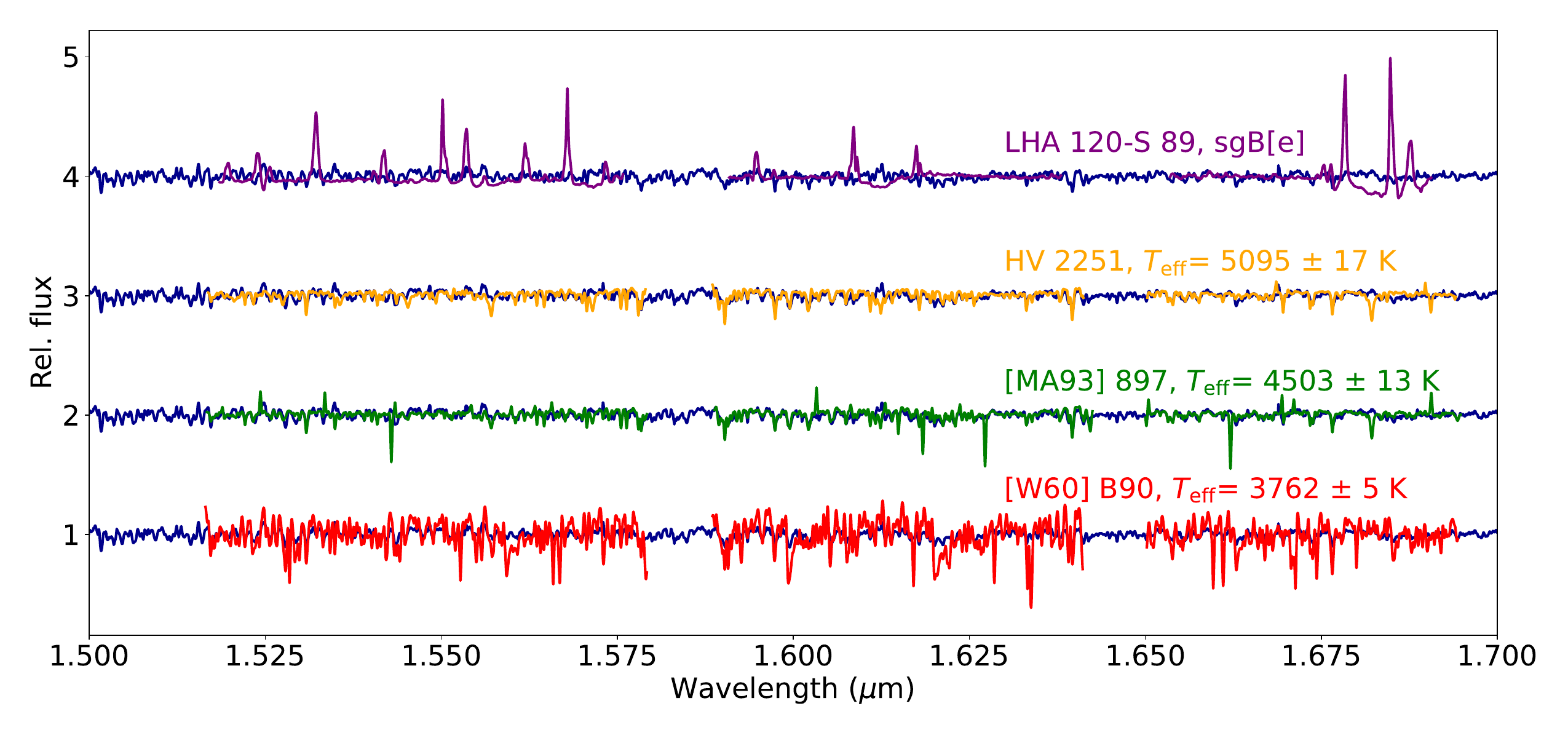}}
        \caption{\textbf{Constraining the \teff~of WOH~G64 in the near-IR}. Comparison of the $H$ band of WOH~G64 from X-Shooter (blue) in 2016  with the APOGEE spectra of four supergiants of the LMC, indicating their \teff~from APOGEE.}
   \label{fig:fig_Hband_comparison}
\end{figure}

We obtained optical spectroscopy of WOH~G64 in 2021 and complemented this with two additional spectra retrieved from the ESO archive (Table~\ref{tab:spec_obs}). Fig.~\ref{fig:fig_spec_comparison} presents an extraordinary change in the optical spectral features accompanied by the dramatic photometric transition in 2014 (see also Extended Data Fig.~\ref{fig:fig_spec_2016_2021_Ha_and_Ca_triplet}). In the UVES spectrum from 2007, WOH~G64 exhibits spectral features characteristic of an M6~Ie-type RSG \cite{Solf1978}, including deep TiO bands and strong Ca~\textsc{ii} triplet absorption lines. Remarkably, we detect strong double-peaked emission from forbidden emission lines (e.g., [N~\textsc{ii}] and [S~\textsc{ii}]), strangely seen in RSGs (Extended Data Fig.~\ref{fig:fig_forbidden}). Multi-epoch spectroscopy after the transition revealed the most extreme change ever seen in the optical spectral features of a RSG. The optical spectral range is dominated by asymmetric emission lines from permitted (e.g., Fe~\textsc{i}, H~\textsc{i}, He~\textsc{i}, Ni~\textsc{i}, and Ti~\textsc{i}) and forbidden transitions (e.g., [Ca~\textsc{ii}], [Fe~\textsc{ii}], [N~\textsc{ii}], [O~\textsc{i}], and [S~\textsc{ii}]) leading to a B[e]-star classification in 2016 and 2021 (see Methods) \cite{Lamers1998, Kraus2019}. Simultaneous near-IR spectroscopy in 2016 revealed a composite spectrum (Extended Data Fig.~\ref{fig:fig_spec_nearIR}). While the $YJ$ bands are dominated by strong H~\textsc{I} and He~\textsc{I} emission, similarly to the optical spectral range, the $H$ band is characterized by neutral atomic absorption lines typically seen in late-G to early-K stars \citep{Rayner2009}. Fig.~\ref{fig:fig_Hband_comparison} presents a comparison of the $H$-band spectrum of WOH~G64 with that of a supergiant B[e] star (sgB[e]), two Yellow Supergiants (YSGs), and a RSG. If WOH~G64 was a sgB[e], its full spectral range would resemble that of a hot B-type star embedded in low-density CSM, typically with ring or disk-like morphology. However, Fig.~\ref{fig:fig_Hband_comparison} shows that the $H$-band spectral features of WOH~G64 closely resemble those of YSGs. Based on this comparison, we assign $\teff = 4,800 \pm 300$~K to WOH~G64, as both YSG spectra reproduce the observed spectral features reasonably well. The absence of typical RSG features in the $H$-band further supports the spectral transition reported in the optical. Therefore, the composite spectral features of a warm star in the near-IR and a B[e] star in the optical indicate that WOH~G64 is a massive symbiotic binary (SymB[e]) \citep{Lamers1998}. 


\section*{Discussion}

\subsection*{Symbiotic Binary}

Symbiotic stars are interacting binaries consisting of a cool (super)giant donor star transferring material to an accreting companion \citep{Allen1984, Mikolajewska2007}. A notable example is VV~Cephei, a massive Galactic symbiotic binary composed of an M2~I red supergiant and a B-type main-sequence star. The B-type companion is surrounded by an accretion disk that produces the SymB[e] spectral features \citep{Cowley1969, Wright1977}. This system has a very wide orbit, with a period exceeding 20 years, which limits the interaction between the two components. However, the nature of a symbiotic binary system depends on its orbital configuration. For instance, D-type symbiotic binaries are strongly interacting systems that contain pulsating red giants enshrouded in thick dust shells. These systems often exhibit complex morphologies, including equatorial rings and bipolar outflows (e.g., R~Aqr) \citep{Liimets2018}.

The SymB[e] scenario for WOH~G64 offers a plausible explanation for its extreme properties, which are difficult to reconcile with a single RSG. WOH~G64 is surrounded by a torus of 3–9~\msun \citep{Ohnaka2008}. The dusty torus may explain the $\sim$850-day periodic photometric variability observed in the 1990s, which is driven primarily by $A_{\rm V}$ changes. This variability could represent half of an orbital period of $\sim$1,700~days and may correspond to symmetric substructures within the torus, such as spiral arms (see Methods). In this scenario, the brightness maximum would occur when the line of sight passes through a cavity, while the minimum would coincide with the passage of a spiral arm across the line of sight. Bipolar outflows ejected by the accretor during mass transfer \citep[e.g.,][]{Soker2015, Dori2023} could account for the double-peaked forbidden emission observed when WOH~G64 was an M-type star (see Methods). The He~\textsc{I} emission detected after the transition might originate in the circumstellar nebula via photoionization by a hot companion \citep[e.g.,][]{Cowley1969}, or in an accretion disk around the companion \cite[e.g.,][]{Beristain2001}. The radial velocity (RV) shifts seen in the Sc~\textsc{ii} and Paschen series absorption lines in 2021 (Extended Data Fig.~\ref{fig:fig_RV}) are also consistent with a binary scenario (see Methods). 

The considerable complexity of the system, likely involving contributions from at least five components: the two stars, their respective winds, and a circumbinary disk, renders spectral disentangling ineffective for determining the physical parameters of either star. First, the absence of atmospheric models appropriate for Yellow Hypergiants and their winds prevents us from extracting accurate stellar parameters for the primary component. Likewise, we cannot reliably constrain the highly variable $A_V$ from the spectral slope, as it is strongly degenerate with the $T_{\rm eff}$ of the primary. Without robust constraints on the primary star or the extinction, it is impossible to estimate the contribution of the secondary star to the observed blue excess. Future spectroscopic monitoring in both the optical and near-IR is key to tracing RV shifts in key spectral lines and to better disentangle the different components. WOH~G64 also shows an elongated central emission extending beyond approximately 9 stellar radii, assuming $R = 1,500$~\rsun, as resolved by $K$-band interferometry \cite{Ohnaka2024}. Consequently, the direct detection of a companion in these observations is not feasible, since it would lie within the core of the emission, where both components are expected to interact and  remain spatially unresolved.

\subsection*{The dramatic transition}

RSGs can temporarily shift in spectral type during dimming events, as observed in HV~11423 and [W60]~B90 \citep{Massey2007, MunozSanchez2024}. However, the spectral and photometric changes in WOH~G64 after its 2014 transition have persisted over time, indicating a more permanent transformation. The absence of any outburst rules out a violent event, such as a merger or the explosion of the RSG, as the cause of the transition. Instead, we adopt a binary scenario, where both components were present before and after the transition. The nearly constant $I$-band brightness across the transition suggests that the same stellar component continued to dominate the system. The disappearance of TiO molecular bands, the brightening of approximately $\Delta V \approx 2$~mag, and the change in the $V-I$ color are consistent with a rise $\Delta T_{\rm eff} > 1,000$~K (Extended Data Fig.\ref{fig:color_MARCS}), likely consequence of the partial loss of the outer stellar envelope. Interferometric $K$-band observations revealed a cocoon around WOH~G64 \citep{Ohnaka2024}. Using the 90~km~s$^{-1}$ outflow velocity inferred from Ca~\textsc{ii} P-Cygni profiles in 2021 (see Methods) and the measured distance to the ring, we estimate a dynamical timescale pointing to early 2013. This temporal coincidence with the transition suggests a possible causal link with the loss of the outer envelope and the cocoon. 

Models on RSGs exploding within dense CSM predict a double-peaked H$\alpha$ emission-line  profile, formed by broad emission with a narrow core, which arises from the electron-scattering from the unshocked CSM at 16 hours after the SN \citep{Dessart2025}. Five days later, a P-Cygni profile appears in the H$\alpha$ profile, with a blue-wing shoulder produced by the fast-moving, dense shell and the unshocked CSM. WOH~G64 shows analogous behaviour (Extended Data Fig.~\ref{fig:fig_spec_2016_2021_Ha_and_Ca_triplet}), but on time scales of years rather than days. While the models assume shell velocities of 7,000~km~s$^{-1}$ for the explosion, adopting 90~km~s$^{-1}$ for the shell ejected by WOH~G64 naturally stretches these changes to several years. Consequently, the post-transition emission-line profiles may also be shaped by interactions between the newly ejected shell and the pre-existing CSM, further complicating spectral disentangling and modelling.

\begin{figure}[H]
   \centering
   \resizebox{0.7\hsize}{!}{\includegraphics{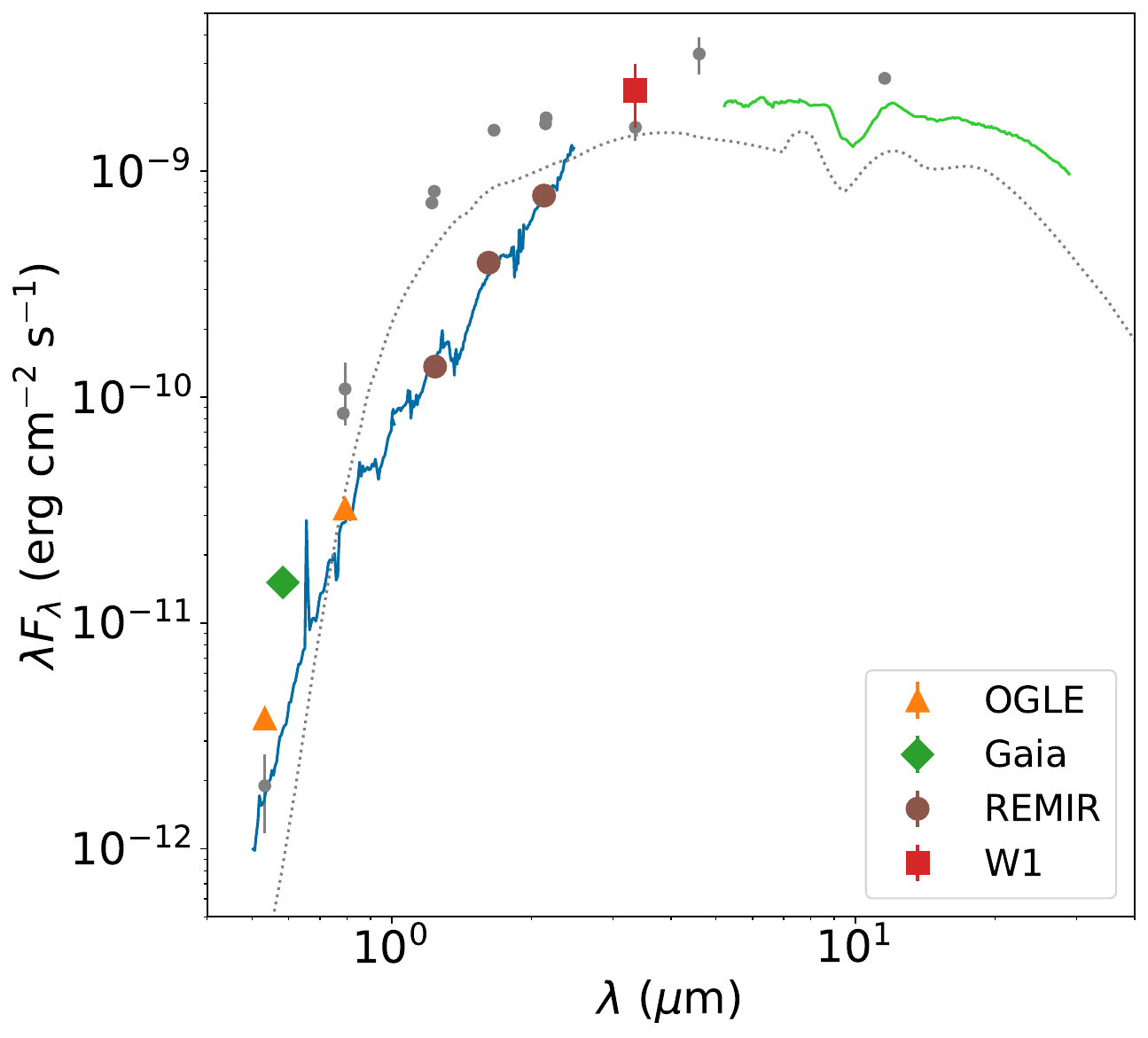}}
        \caption{\textbf{Evolution of the SED of WOH~G64 before and after the transition.} The gray points indicate the photometry before the transition \cite{Antoniadis2024}, while the colored symbols indicate the photometry after the transition (Methods). The blue line is the X-Shooter spectrum, the green line is the IRS spectrum in green \cite{Houck2004}, and the dotted line is the best-fit model before the transition \cite{Antoniadis2024} scaled to $\log(L/\lsun)=5.45$ \citep{Ohnaka2008}}.
   \label{fig:fig_SED}
\end{figure}

\begin{figure}[H]
    \centering
    \resizebox{0.8\hsize}{!}{\includegraphics{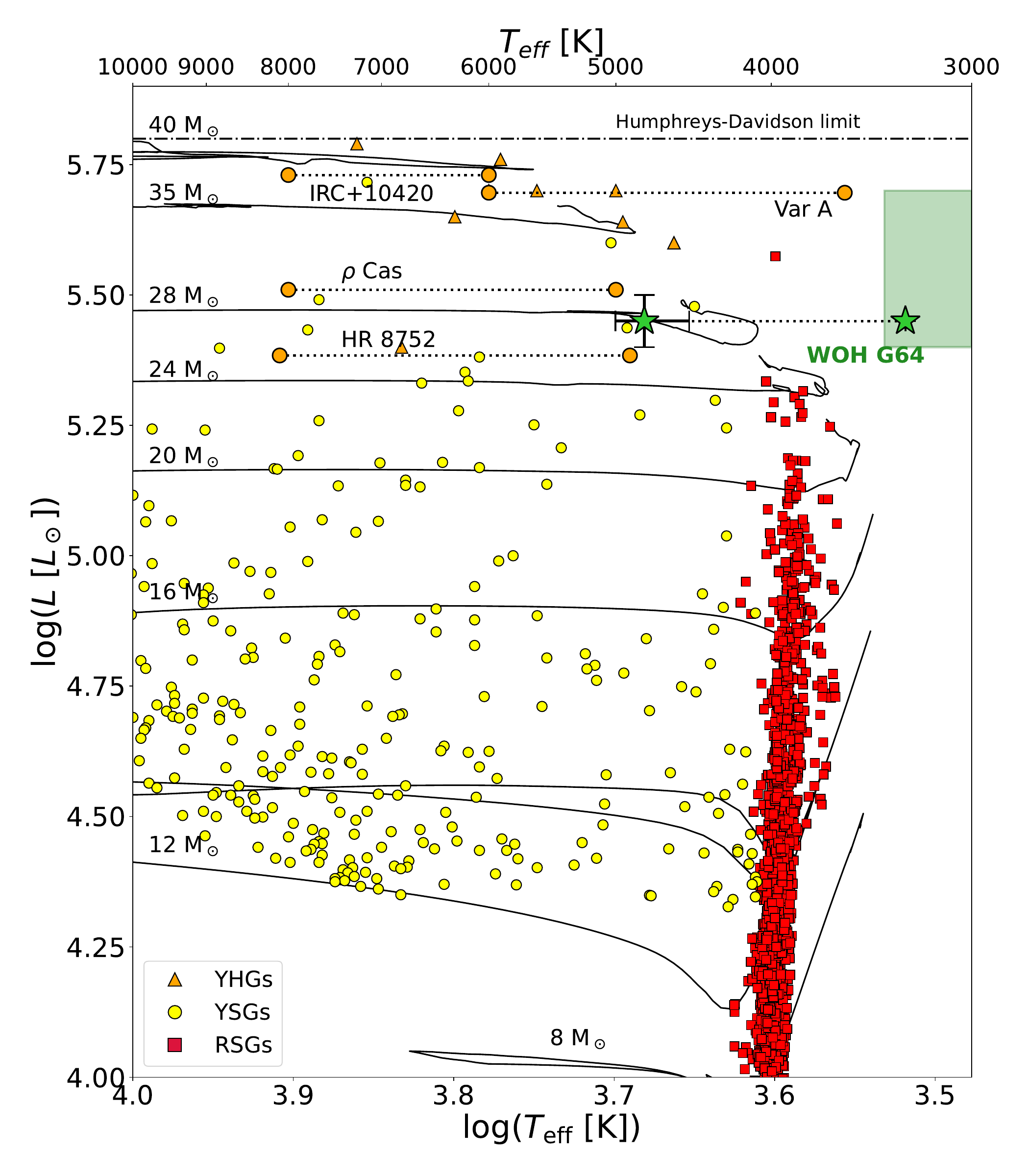}}
        \caption{\textbf{Hertzsprung-Russell diagram showing the transition of WOH~G64}. Other extreme YHGs transitions (orange dots) and the LMC population of YHGs \citep[orange triangles,][]{Kourniotis2022, Humphreys2023}, YSGs \citep[yellow circles,][]{Neugent2012}, and RSGs \citep[red squares,][]{Antoniadis2024} are shown for comparison. \textsc{mist} evolutionary tracks are presented in black \cite{Dotter2016, Choi2016}. The green shaded area represents the range of parameters derived for WOH~G64 prior to this work \cite{Levesque2009}. }
   \label{fig:fig_HRD}
\end{figure}

The observed increase in $T_{\rm eff}$ implies that WOH~G64 evolved from a late-M RSG to a late-G or early-K YHG. Its spectral energy distribution (SED) in 2016 shows a brightening in the $V$-band and dimming in the near-IR, consistent with a warmer temperature (Fig.~\ref{fig:fig_SED}). Luminosity estimates assuming spherical symmetry yield $\log(L/L_\odot) = 5.75 \pm 0.01$ before and $5.57 \pm 0.01$ after the transition (see Methods). The small error bars reflect only the uncertainties in the photometric data and in the distance to the LMC when integrating the SEDs. However, the true uncertainty might be larger, up to $\sim$0.2--0.3~dex, due to additional factors such as variations in scattered light and the intrinsic variability of the system. We finally adopt $\log(L/L_\odot) = 5.45 \pm 0.05$ from models accounting for the nonspherical CSM geometry of WOH~G64 \citep{Ohnaka2008}. The updated position of WOH~G64 on the Hertzsprung–Russell diagram (Fig.~\ref{fig:fig_HRD}) aligns more closely with the evolutionary track of a star with initial mass 28~$ M_\odot$, rather than the 25~$M_\odot$ track previously suggested \citep{Ohnaka2008}. The inferred current radius is approximately 800~$R_\odot$, about half the previously estimated size of 1540~$R_\odot$ \citep{Levesque2007}. These values should be taken as indicative rather than definitive, given the significant uncertainties involved in accurately determining the luminosity of such highly obscured evolved systems \citep[e.g.,][]{Levesque2009}.

\subsubsection*{Common envelope evolution}

Binary interactions in a symbiotic system provide a compelling framework to explain the observed properties of WOH~G64, and may have also driven the dramatic transition. Mass transfer is expected for RSGs in wide binaries, potentially leading to CE evolution depending on the system’s properties \citep{Paczynski1972, Ercolino2024}. However, predicting the outcome of the CE phase remains challenging, as it strongly depends on the stellar parameters and is a multi-timescale, complex, 3D process, computationally difficult to model, with many uncertainties  \citep{Ivanova2013}. WOH~G64 has always shown an unusually cool and extended atmosphere compared to canonical RSGs \cite{Levesque2017}, which may, in fact, reflect the presence of an extended CE surrounding the binary. During the CE phase, the system may enter a slow, self-regulated phase that can last hundreds of years. In this regime, the energy from orbital decay is efficiently transported to the surface and radiated away \cite{Clayton2017}, resulting in large-amplitude, repeating pulsations. These pulsations can eject shells every few decades, producing a time-averaged mass-loss rate of \Mdot~$\sim 10^{-3}$ \msun~yr$^{-1}$. This scenario is consistent with both the high mass-loss rate inferred for WOH~G64 (\Mdot~$>10^{-4}$ \msun~yr$^{-1}$; \cite{Antoniadis2024}) and the pulsation-like loop observed in the dimming event before the transition (Extended Data Fig.~\ref{fig:color_MARCS}). 

A short mass-loss episode of the order of a solar mass is physically plausible within the framework of CE evolution. While the full CE interaction can last $\sim$1000 years on the thermal timescale of the donor star, much faster phases are expected near the termination of the self-regulating spiral-in \citep{Ivanova2013}. Numerical work supports this, including recombination energy leading to complete ejection in simulations \citep{Nandez2015}, and recent 3D calculations show that up to 80\% of a red giant's envelope can be expelled in just $\sim$400 days, with full unbinding after $\sim$1500 days \citep{Valsan2023}. Such results point to envelope-stripping on the order of a year being feasible, especially if the outer layers are loosely bound or already destabilized by prior mass transfer. In this scenario, a RSG could undergo a slow spiral-in while still appearing red, followed by a brief, dynamical mass-loss episode that removes part of the envelope, leaving behind a temporary yellow transient object.

\subsubsection*{Single-stellar physics: an outstanding eruption}

Beyond binary interactions within the symbiotic system, the transition could be driven by single-stellar physics. The luminosity of WOH~G64, $\log(L/\lsun)=5.45 \pm 0.05$ \cite{Ohnaka2008}, is close the Humphreys-Davidson limit, $\log(L/\lsun)\sim5.5-5.8$ \cite{Humphreys1979, Davies2020_luminosity} where extreme physical processes associated with significant eruptions are expected \cite{Glatzel1993}. 
Var~A, one of the most luminous stars in M33 \citep{Hubble1953}, underwent an eruption lasting over 45~yr and serves as a notable example. Eruptions in other YHGs are typically recurrent but short-lived \cite{Koumpia2020, vanGenderen2025}, rather than several decades. During the eruption, Var~A exhibited late-M spectral features due to its optically thick wind ($\dot{M}>10^{-4}~\msun~\rm yr^{-1}$) \cite{Humphreys1987, Humphreys2006}. When the eruption ended, Var~A returned to its YHG state, exhibiting F-type features along with both forbidden and permitted emission lines. Similar to Var~A, WOH~G64 might have been a YHG undergoing an eruption for more than 30~yr, until it returned to a quiescent state in 2014 when the false pseudo-atmosphere dissipated. Fig.~\ref{fig:fig_HRD} compares the evolutionary transitions of WOH~G64 and Var~A, along with other well-known YHGs. The apparent higher luminosity of Var~A might be overestimated, just as in WOH~G64, where direct integration of its SED yields $\log(L/\lsun)=5.75$ (see Methods). Var~A also exhibited a decrease in integrated luminosity after returning to quiescence, explained due to the radiation escaping through low-density regions or holes in the complex CSM rather than along the direct line of sight \cite{Humphreys2006}. Unlike WOH~G64, the eruption and subsequent transition of Var~A are poorly constrained. During the 45~yr eruption period, Var~A was photometrically observed only once between 1955 and 1982, leaving 60\% of the eruption duration undocumented. The duration of the transition from a late-M to an F-type spectrum is only constrained by an upper limit of 5 years. Additionally, the post-transition $\teff$ of Var~A was estimated based on its optical spectral features. Following the same approach in WOH~G64, the \teff~increase would then correspond to a change from 3200~K \citep{Ohnaka2008} to $\gtrsim$~20000~K. SED modeling of Var A with a dusty torus yielded low extinction ($A_V = 0.6$~mag) \cite{Humphreys1987}, in contrast to the extreme extinction in WOH~G64 ($A_V > 6$~mag) \cite{Levesque2009}. The discrepancy in $A_V$ between the two objects could be explained by the differing inclination angles of their respective systems. The dramatic optical changes in WOH~G64, its fast transition, and high obscuration suggest a more extreme environment than that of Var~A. 


\subsection*{Future of the system}

Our findings challenge current models of massive star evolution by underscoring the critical role of binary interactions and episodic eruptions in shaping the fate of evolved massive stars. In particular, WOH~G64 offers valuable insight into the late evolutionary stages of the most luminous and heavily mass-losing RSGs, and the long-standing ``RSG problem." Recent studies of extreme RSGs such as VY~CMa and NML~Cyg have uncovered potential evidence of companions, yet the mechanisms driving their episodic mass-loss events remain poorly understood \citep{Smith2025, deBeck2025}. Determining whether the extreme properties of these RSGs \citep{Jones2025} arise from intrinsic stellar processes or from binary interaction is essential for constraining the evolution of very luminous RSGs \citep{Marchant2024}.


We cannot predict the future of WOH~G64 due to the poorly constrained physical and orbital parameters, and because we do not know whether single-stellar physics or binary interactions drive the system evolution. Observations obtained in 2025 \cite{vanLoon2026} support our interpretation of WOH~G64 as a binary system and further demonstrate that the system is affected by variable dust reddening. However, WOH~G64 currently exhibits optical features consistent with a Red Supergiant, rather than the B[e] characteristics observed during 2016–2021. This transition back to an RSG can be explained as a return to an optically thick wind and the beginning of another eruption under the single-star physics scenario. Predicting the future and recurrence of such eruptions remains uncertain because the driving mechanism is still under debate. Meanwhile, within the CE scenario, the emergence of new cool-star signatures may be explained by the formation of a new, extended cool envelope after the ejection of the outer layers during the 2013--2014 event. The present orbital configuration will determine whether interaction continues, potentially leading to subsequent phases of mass transfer or recurrent CE ejections \cite{Clayton2017}, as seen during the 2014 episode.

With the advent of modern surveys that continuously monitor the sky, the early detection of SN explosions and subsequent rapid spectroscopy have revealed the presence of pre-SN CSM surrounding Type II SN progenitors. One proposed explanation is a major outburst occurring within the final year before the explosion \citep{Davies2022}, while other authors favor the pulsation-driven pre-SN wind scenario decades before the explosion \citep[e.g.,][]{Sengupta2025}. Late carbon-burning pulsations in RSGs \citep{Bronner2025, Laplace2025} can produce variations of up to 1~dex in $\log(L)$ and $>2{,}000$~K in $\teff$ decades before the explosion. However, the post-transition behaviour of WOH~G64 does not follow this periodic pattern, remaining stable from 2014–2019, thereby ruling out pre-SN wind pulsations as the mechanism behind the eruption. Nonetheless, the YHG will likely end its life first, and binarity together with CSM interaction could strongly shape the resulting SN type. For wide initial periods such as those expected in WOH~G64, the outcome is likely a Type IIP/L, IIb, or IIn \citep{Ercolino2024, Ercolino2025}, depending on the remaining envelope mass and CSM density at core collapse \citep{VanDyk2002, Folatelli2015, Smith2014}.

WOH~G64 is embedded in a dense, strongly asymmetric CSM, showing elongated $K$-band emission of $\sim 1 \times 10^{15}$~cm that matches the inner radius of the dusty torus, which extends to $\sim 3 \times 10^{16}$~cm \citep{Ohnaka2008, Ohnaka2024}. Ejecta expanding at $10,000$~km s$^{-1}$ would reach these radii at $\sim$10 and $\sim$340 days after explosion. The strong CSM asymmetry implies that the SN observables will depend on inclination: expansion through the equatorial plane would produce strong interaction with the dense disk-like structure, whereas expansion along the polar axis would proceed without obstruction, yielding weak or absent interaction signatures. Intermediate inclinations would naturally produce intermediate behaviors. The dense CSM also obscures the central star, making it appear redder even though its intrinsic temperature is hotter than that of a typical RSG. This underscores that some SN progenitors may be misclassified as RSGs due to heavy CSM obscuration when relying solely on photometric colours.

As an alternative to explosion, the star may undergo direct collapse into a black hole \citep{OConnor2011}, or merge with its companion as a result of continued binary interaction \citep{Sana2012}. WOH~G64 thus provides critical insight into post-RSG evolution and the formation of dense circumstellar environments seen in core-collapse SNe. Continued spectroscopic and photometric monitoring of WOH~G64 will be essential to constrain its binary properties, uncover the mechanism behind its dramatic transition, and predict its ultimate fate.

\newpage

\section*{Methods}\label{methods}

\subsection*{Photometric catalogs}\label{sec:phot_catalogs}

We assembled the photometry of WOH~G64 available over the past three decades from: ALLWISE \citep{WISE2010, ALLWISE2014}, ATLAS \citep{ATLASmain2018}, \textit{Gaia} DR3 \citep{Gaiamision2016, GaiaDR32023}, the MACHO project \citep{MACHO1997}, NEOWISE-R \citep{NEOWISE2011}, and OGLE \citep{Udalski1997,OGLEIII2008, Udalski2015}. We applied the following criteria to each survey to select the most reliable data.  

\begin{itemize}
    \item ATLAS: We chose the reduced images option in the forced photometry server to obtain the light curve \citep{ATLASvariable2018, ATLASserver2021}. We used data points with flag $err=0$, $chi/N<100$, and error lower than 0.2~mag. We binned the measurements within the same night to decrease the scatter. 
    
    \item MACHO: We used the approximation $X \approx X_{M,t}+z_0+2.5\log(ET)$ of Eq. 1 and 2 in \cite{MACHOcalibration1999} to produce the $V_{KC}$ and $R_{KC}$ photometry. In our approximation, $X$ refers to the $V_{KC}$ and $R_{KC}$ bands, $z_0$ is the corresponding photometric zero point, and $ET$ is the exposure time. We keep measurements with errors lower than 0.1~mag and discard outliers with small errors by applying a 3$\sigma$ clipping. 

    \item NEOWISE-R: WOH~G64 has been observed on 21 main epochs since the mission started in 2014. We binned all the measurements within two consecutive weeks by taking the median value and the uncertainty of the median. We discarded the data with \textit{qual\_frame}$=0$ and $chi^2>20$. NEOWISE-R photometry differs from ALLWISE for targets brighter than $W1<8$~mag. We applied the offsets provided in the supplementary material of the NEOWISE-R mission. $W2$ data is severely saturated.   

    \item OGLE: We used the V and I-band data from OGLE-III and OGLE-IV databases.

\end{itemize}

\subsection*{Spectroscopic Observations}\label{sec:longslit_obs}

We obtained optical spectroscopy of WOH~G64 with the Magellan Echellette (MagE) spectrograph \citep{MagE2008} on the 6.5-m Baade telescope at Las Campanas Observatory, Chile. We used the $1.0" \times 10"$ long-slit, providing a wavelength coverage of $3,500-9,500$~\r{A}, a spectral resolution of $R\sim4,000$, and a spatial resolution of 0.3$"$~px$^{-1}$ with binning $1\times1$. We used the MagE pipeline \citep{Kelson2000, Kelson2003} for the bias and flat correction. We manually continued the reduction using the \textsc{echelle} package of IRAF. Finally, we used the flux standard to calibrate the spectrum with the IRAF routines {\sc standard} and {\sc sensfunc}.

We further collected two spectra from the ESO archive, obtained with the instruments UVES \citep{Dekker2000} and X-Shooter \citep{Vernet2011} on the UT2 at the Very Large Telescope, Cerro Paranal, Chile. The UVES observations took place under the program ID 080.D-0508(A) (PI: A. Manchado). Two long exposures of 1800~s and a short exposure of 300~s were obtained with the Cross Disperser \#4, covering the spectral range $5,655-9,612$~\r{A}, providing a spectral resolution $R \sim 57,000$. The public data are not flux calibrated and were reduced using the official UVES pipeline v5.10.13 \citep{Ballester2000}. The X-Shooter spectra were obtained on the program ID 097.D-0605 (PI: S. Goldman) with the echelle slit spectroscopy mode. The 1.3" slit was used for the UVB arm, providing $R \sim 4100$, while the 1.2" slit was used for the Vis and NIR arms, yielding $R \sim 6500$ and $R \sim 4,300$ respectively. The spectral coverage was $2,989-5,560$~\r{A}, $5336-10,200$~\r{A}, and $0.99-2.48$~$\rm \mu m$ for the UVB, VIS, and NIR arms. Four images were combined to produce the final spectrum, with a total integration time per pixel of 980s in the UVB and NIR arms, and 624s in the VIS arm. The data were reduced with the official X-Shooter pipeline v2.7.0b \citep{Modigliani2010}. We corrected the final NIR spectrum for telluric absorption with \textsc{molecfit} v4.3.1 \cite{Smette2015}, which calculates a synthetic spectrum of the Earth’s atmosphere based on local weather conditions and standard atmospheric profiles. X-Shooter is a supported instrument by \textsc{molecfit}, hence we used the default parameters set up in the workflow. We refined the spectral regions considered for the telluric correction to optimize the cleaning in the $J$, $H$, and $K$ bands. 

Table~\ref{tab:spec_obs} shows the log of the spectroscopic observations, listing the telescope, instrument, UT date of the observation, exposure time, the slit width, and $R$.

\subsection*{Spectral Energy Distribution of WOH~G64}\label{sec:SED}

We investigated the differences in the spectral energy distribution (SED) of WOH~G64 before \citep{Antoniadis2024} and after the transition (Fig.~\ref{fig:fig_SED}). We used the X-Shooter observations from July 2016 as a reference to construct the new SED. We collected OGLE $V$ and $I$, W1, and $G$ from the closest day to the X-Shooter observation. The shape of the SED of WOH~G64 at 10~$\mu$m was demonstrated to be constant since 2004
\citep{Ohnaka2024}. Hence, we added the Spitzer InfraRed Spectrograph (IRS) Enhanced Product \citep{Houck2004} as a reference in the mid-IR. We also used the $JHK'$ photometry from 2024 \citep{Ohnaka2024} from the IR camera on the Rapid Eye Mount telescope (REMIR) \cite{Calzoletti2005}, as it was found to match the SED of the X-Shooter spectrum. 

We attempted to fit the post-transition SED using the fitting procedure and assumptions described in \citep{Antoniadis2024}, but without success. The new SED is slightly brighter in the optical, significantly fainter in the near-IR, and nearly unchanged in the mid-IR. Our models could not reproduce this behavior, as they failed to match the optical slope when optimizing the fit in the near-IR, and vice versa. This failure may arise from several factors: the assumption of spherical symmetry, which is far from being valid for WOH~G64 \citep{Ohnaka2008, Ohnaka2024}; the presence of multiple components contributing to the SED, such as the companion; or the thermal emission from the layer expelled during the transition.

We integrated the luminosity assuming spherical symmetry, yielding $\log(L/L_\odot) = 5.75 \pm 0.01$ before and $5.57 \pm 0.01$ after the transition. These uncertainties are derived from the propagation error from the photometric data and the distance to the LMC.  The true uncertainty may be up to $\sim$0.2–0.3~dex larger, due to additional factors such as variations in scattered light and the intrinsic variability of the system. 

\subsection*{Causes of the optical variability and color change}\label{sec:color_change}


When WOH~G64 was classified as a late M-type RSG, it was also identified as a Mira variable based on its large photometric amplitude and semi-regular periodicity \citep{Soszynski2009}. Mira variables typically exhibit cyclic loops in a color–magnitude diagram over a pulsation period, driven by intrinsic changes in stellar radius and \teff~\citep{Boyd2021}. However, the color–magnitude diagram of WOH~G64 derived from MACHO photometry does not exhibit this cyclic pattern (Extended Data Fig.~\ref{fig:color_MARCS}); instead, it follows an almost linear trend. We used \texttt{marcs} models \citep{Gustafsson2008} and applied reddening to produce synthetic photometry and to test if variations $A_V$ replicate the linear trend. The only published estimate for WOH~G64 during the 1990s is $\teff=3000$~K \citep{vanLoon2005}; however, we adopted \teff~$= 3300$~K, the lower limit of the \texttt{marcs} grid. In the absence of official MACHO transmission curves, we downloaded the Kron–Cousins filter profiles of the SOAR Adaptive Module (SAM) at Cerro Tololo Inter-American Observatory (CTIO) from the Spanish Virtual Observatory (SVO) repository. Our synthetic photometry shows that the observed linear trend in the color-magnitude diagram can be reproduced if extinction dominates the photometric variability. For \teff~$= 3300$~K, matching the observed colors requires periodic extinction variations of $\Delta A_V \sim 3$ mag, ranging from $A_V = 7.7$~mag at maximum brightness to $A_V = 10.9$~mag at minimum. These absolute values are model-dependent and should be interpreted with caution, as the real \teff~of WOH~G64 during the 1990s may be cooler. A cooler temperature would imply correspondingly lower values of $A_V$ needed to reproduce the observed photometry. These results imply that $A_V$ dominant factor causing the variability in the MACHO light curve.

A different behavior emerges when the analysis is repeated using the OGLE $V$ and $I$ bands between 2010 and 2017. During the dimming event leading up to the transition, WOH~G64 traces a loop in the color–magnitude diagram, indicating intrinsic changes in the system beyond extinction alone. Furthermore, the abrupt color “jump” observed in Extended Data Fig.~\ref{fig:color_MARCS} from before to after the transition can only be explained by a temperature increase of $\Delta \teff > 1,000$~K, assuming a moderate extinction of $A_V \sim 4.0$~mag. Although absolute values remain uncertain due to poorly constrained physical conditions during the transition, the large color change, despite the $I$-band brightness remaining nearly constant, strongly supports a significant increase in \teff.

\subsection*{Mid-infrared light curve} \label{sec:mid-IR_lightcurve}

We investigated the mid-infrared (mid-IR) variability of WOH~G64 using W1-band data from NEOWISE-R (Fig.~\ref{fig:fig_lightcurve}). The start of the mission in 2014 coincides with the onset of the dramatic transition observed in WOH~G64. However, we do not detect any significant variation when compared to WISE data from 2010, before the transition. The W1 photometry remains generally constant at $\rm{W1} \approx 5.4$~mag, with a variability amplitude of only $\Delta\rm{W1} = 0.1$~mag, which is smaller than the uncertainty. We report three brightenings, each with an amplitude of $\Delta\rm{W1} \approx 1$~mag, occurring in mid-2019, mid-2021, and mid-2023 (MJD = 58,631, 59,359, and 60,094, respectively), suggesting a possible recurrence timescale of $\sim$750 days. Due to the low photometric cadence, the exact timing of the peaks is uncertain. Interestingly, the first mid-IR brightening coincides with a peak in the optical light curve. No optical data are available for the second brightening, while the third mid-IR peak is followed by a delayed optical brightening. These events may be linked to interactions with the disk-like structure surrounding WOH~G64, similar to the $\sim$1 mag mid-IR outbursts observed in some Be stars, when their circumstellar disks build up and dissipate \citep{Jian2024}.

\subsection*{Optical spectral classification}\label{sec:opt_spectroscopy_methods}

The 2007 unpublished UVES spectrum shows characteristic features of late M-type RSGs, such as deep TiO bands, strong Ca~\textsc{ii} triplet, and absorption lines from Fe~\textsc{i} and Ti~\textsc{i} (Fig.~\ref{fig:fig_spec_comparison} \& Extended Data Fig~\ref{fig:fig_spec_2016_2021_Ha_and_Ca_triplet}). Strangely, only one Mg~\textsc{i} line from the $\lambda\lambda$8500-8800 series \citep{Tabernero2018} is present. We followed the spectral classification based on the strength of TiO and VO bands \cite{Solf1978}. The existence of the heads of TiO at $\lambda\lambda$8199, 8506, 8515 suggests that its spectral type is M5 or later. We classify it as an M6 star because of the existence of TiO at $\lambda\lambda$8206, 8251, 8268, 8373, and 8420 and the strength of the TiO bands at $\lambda\lambda$8432, 8442 compared to the Ca~\textsc{ii} lines. We only identify the VO head at $\lambda$7865 out of the two that start to appear 
at M6 ($\lambda\lambda$7865, 7897). We also detect emission from H$\alpha$ as well as forbidden lines, such as [N~\textsc{i}] $\lambda$5198, [N~\textsc{ii}] $\lambda\lambda$6548 and 6583, [S~\textsc{ii}] $\lambda\lambda$6716 and 6731, [O~\textsc{i}] $\lambda\lambda$6300 and 6364. We highlight the double-peaked [N~\textsc{ii}] and [S~\textsc{ii}] emission (Extended Data Fig.~\ref{fig:fig_forbidden}), with their components separated by $\sim$210~km~s$^{-1}$ and $\sim$100~km~s$^{-1}$, respectively.

The X-shooter spectrum from 2016 reveals dramatic changes (Fig.~\ref{fig:fig_spec_comparison} and Extended Data Fig.~\ref{fig:fig_spec_2016_2021_Ha_and_Ca_triplet}). The TiO bands have completely disappeared, and double-peaked and asymmetric emission lines dominate the spectrum. The only absorption lines in the optical range are K~\textsc{i} $\lambda\lambda$7665 and 7699, and Na~\textsc{i} $\lambda\lambda$5890 and 5896. We detect the same forbidden transitions as in 2007, plus several  [Fe~\textsc{ii}] lines and the [Ca~\textsc{ii}] $\lambda\lambda$7291 and 7324 doublet (Extended Data Fig.~\ref{fig:fig_forbidden}). The majority of permitted transitions arise from Fe~\textsc{i}, Fe~\textsc{ii}, Ni~\textsc{i} and Ti~\textsc{i}. Additionally, the Ca~\textsc{ii} triplet and H$\alpha$ are double-peaked, while the Paschen series and He~\textsc{i} $\lambda$6678 show asymmetric emission. Notably, the O~\textsc{i} triplet $\lambda$8446 is in emission, while the O~\textsc{i} triplet $\lambda$7774 is barely in absorption. 

The MagE spectrum obtained in 2021 reveals various changes with respect to 2016. The asymmetric and double-peak emission lines still dominate the optical range, but new absorption lines have emerged. Along with the K~\textsc{i} and  Na~\textsc{i}, Si~\textsc{ii} $\lambda\lambda$6347 and 6371, Sc~\textsc{ii} series at $\lambda\lambda$ 5500-5700, the O~\textsc{i} triplets $\lambda\lambda$8446 and 7774, and Paschen series also appear in absorption. Moreover, H$\alpha$ and the Ca~\textsc{ii} triplet have P Cygni profiles, rather than the double-peaked emission in 2016. We followed the approach by \citep{Humphreys2014} and measured the outflow velocities in H$\alpha$ and the Ca~\textsc{ii} triplet from the minima in their P Cygni profiles relative to the emission line peak, finding $80\pm10$ and $90\pm10$~km~s$^{-1}$, respectively. These outflow velocities are lower than those typically found in the Luminous Blue Variables, Warm Hypergiants, and other Supergiants in M33 and M31 \citep{Humphreys2014}. The change from double-peaked to P~Cygni profiles in the Ca~\textsc{ii}, and from emission to absorption in the O~\textsc{i} triplets and the Paschen series might indicate a softening in the stellar wind intensity. Nevertheless, the emission of permitted transition from Fe~\textsc{i}, Fe~\textsc{ii}, Ni~\textsc{i}, and Ti~\textsc{i}, and all the forbidden transitions remain as in 2016 (Extended Data Fig~\ref{fig:fig_forbidden}).

Extended Data Table~\ref{tab:spectral_lines} lists the identified lines in the optical range of the spectra presented in this work.

\subsection*{Near-IR spectroscopy}\label{sec:spectroscopy_methods}

We show the telluric-cleaned spectrum obtained with the NIR arm of X-Shooter in Extended Data Fig.~\ref{fig:fig_spec_nearIR}, covering the bands $YJ$ (1.0-1.35~$\mu$m), $H$ (1.4-1.75~$\mu$m), and $K$ (2.02-2.45~$\mu$m). We report the main spectral lines in Extended Data Table~\ref{tab:spectral_lines_nearIR}. The strong asymmetric emission from the Paschen series and the He~\textsc{i} triplet at 1.083~$\mu$m dominate the spectrum in the $YJ$ band. We identified various emission series from Ti~\textsc{i} at 1.03-1.08~$\mu$m and 1.26-1.31~$\mu$m and Fe~\textsc{i} at 1.16-1.20~$\mu$m. We also report the presence of the forbidden multiplet of [N~\textsc{i}] at 1.04~$\mu$m, and several absorption lines of Si~\textsc{i}. We note the absence of the Mg~\textsc{i} lines typically found at 1.16-1.20~$\mu$m \citep{Bergemann2015}, which might be filled with emission. The $K$ band shows strong asymmetric emission from Br$\rm \gamma$, a series of Ti~\textsc{i}, and several unidentified emission lines. The spectral range 2.25-2.45~$\mu$m is usually dominated by CO-bands in sgB[e] and cool stars \citep{Rayner2009, Kraus2019}. However, we do not detect well-defined CO-bands, only broad features that might be reduction artifacts; hence, we do not consider them further. 

Comparison of absorption lines in the $H$ band with the near-IR atlas by \citep{Rayner2009} constrains the spectral type of WOH~G64 in 2016. The lack of the Br series indicates a late-G spectral type or later. We observe the metallic forest of atomic absorption lines (e.g., Fe, Si, and Ti) in the spectra of K-type stars (Extended Data Fig.~\ref{fig:fig_spec_nearIR}). The absence of strong molecular CO features and the bump due to H$^{-}$ opacity agrees with an early K-type star. Therefore, we conclude that WOH~G64 has a late-G or early-K spectral type. We also used APOGEE~DR17 \citep{Apogee2022} to compare WOH~G64 with the $H$-band spectra of four stars from the LMC to constrain the \teff. We chose: LHA 120-S 89, a sgB[e] with similar luminosity \citep{Kraus2019} to WOH~G64; [W60]~B90, an M-type extreme RSG undergoing episodic mass loss \citep{deWit2023, MunozSanchez2024}, and two warm supergiants with the high S/N, $\log(g)<1$~dex and $4500<\teff<5500$~K. We selected this \teff~range because the H~\textsc{i} lines become stronger at warmer \teff~and are not present in WOH~G64. The two warm stars are [MA93]~897, a RSG with a hot companion \citep{Patrick2022}, and HV~2251, a yellow supergiant candidate \citep{Neugent2012}. We compare WOH~G64 with the four selected stars in Fig.~\ref{fig:fig_Hband_comparison}, highlighting the \teff~reported in APOGEE. Despite the B[e] features found in the optical regime, this spectral classification is incompatible with the $H$-band. Moreover, WOH~G64 does not show the same RSG features as [W60]~B90, indicating a warmer \teff. However, HV~2251 and [MA93]~897 are similar to WOH~G64. We, therefore, estimate $\teff = 4800 \pm 300$~K for WOH~G64, which agrees with the late-G and early-K classification and the lack of TiO bands in the optical range. However, it cannot explain the He~\textsc{i} emission in the optical and the $J$-band, hinting at a composite spectrum as the result of a binary system. 

\subsection*{Analysis of the forbidden emission}\label{sec:forbidden_emission}


Nebular emission adds crucial clues about the nature of WOH~G64. This emission has been observed in WOH~G64 since its first optical spectrum in the 1980s \citep{Elias1986}. WOH~G64 is not associated with any H~\textsc{ii} region, and no extended emission is detected in the 2D MagE spectral image, indicating that the emission originates from its circumstellar environment. Extended Data Fig.~\ref{fig:fig_RV} shows the heliocentric RV evolution between epochs of the main forbidden and emission lines. Even during the RSG phase, the [N~\textsc{ii}] and [S~\textsc{ii}] lines shows a remarkable peak-to-peak RV separation of $\sim$210 km s$^{-1}$ and $\sim$100 km s$^{-1}$, respectively. A Keplerian disk origin is unlikely: such rotational velocities at the surface of a 1,500~\rsun\ RSG would imply a central mass of $\sim$90~\msun, much higher than the estimated initial mass 25--28~\msun\ for WOH~G64 \citep{Ohnaka2008}. A disk at even larger radii would require a greater central mass, further disfavoring this scenario.

Instead, we attribute the double-peaked forbidden emission lines to bipolar outflows. During its RSG phase, WOH~G64 exhibited enhanced [N~\textsc{ii}]/H$\alpha$ ratios, consistent with N-enriched material ejected and processed through stellar evolution. The high [S~\textsc{ii}]/H$\alpha$ ratio suggests shock excitation as the dominant ionization source \citep{Mathewson1973}. Shocks within these outflows likely produce the observed emission. The [N~\textsc{ii}], [S~\textsc{ii}], and [O~\textsc{i}] lines trace different regions within the outflows, as suggested by their differing peak-to-peak RV separations. These differences might correspond to the ionization potentials of the emitting species: [N~\textsc{ii}] arises from higher-energy regions, followed by [S~\textsc{ii}], and then [O~\textsc{i}], consistent with the trend observed in velocity space. After the transition, the double-peaked emission lines persist, but the [S~\textsc{ii}]/H$\alpha$ ratio decreases significantly. This suggests the emergence of an additional photoionization component, likely from a hot companion, since a warm star with $\teff = 4800 \pm 300$~K is not sufficiently hot to photoionize the CSM.



\subsection*{Radial velocity changes}\label{sec:radial_velocity}


We measured the heliocentric RVs from both emission and absorption lines across three epochs using the IRAF \textsc{splot} task. Extended Data Fig.~\ref{fig:fig_RV} shows the RV evolution of both emission and absorption lines over time.  We also include archival RV measurements from the Ca~\textsc{ii} triplet and the redshifted components of [N~\textsc{ii}] and [S~\textsc{ii}] from previous studies \citep{VanLoon1998, Levesque2009}. For the $H$-band absorption lines in the X-Shooter spectrum, we performed a cross-correlation with a \texttt{marcs} model spectrum of $\teff = 4500$~K and $\log g = 0.0$~dex. Uncertainties are based on the spectral resolution of each instrument unless asymmetriies in the line profiles limited the precision of RV determinations, increasing the associated uncertainties.

The RVs of forbidden lines remain consistent across epochs, supporting an origin in shocks produced by bipolar outflows. Such perpendicular outflows to the orbital plane would not be affected by orbital motion. In contrast, permitted lines display clear variability, both across epochs and between different species within the same epoch, indicating a more complex scenario than a single star. The Ca~\textsc{ii} triplet, typically a reliable RV tracer, was unsuitable in the 2016 and 2021 data due to asymmetric emission and P-Cygni profiles. However, the UVES spectrum yielded an RV of $285 \pm 2$~km~s$^{-1}$, consistent with previous estimates: $292 \pm 15$~km~s$^{-1}$ \citep{Levesque2009} and $\sim300$~km~s$^{-1}$ \citep{VanLoon1998}. Given the similar spectral resolutions, the small differences among these values are not significant and may reflect atmospheric convection, which can cause stochastic shifts of $\sim$10~km~s$^{-1}$ \citep{Kravchenko2019, Kravchenko2021}.

The RVs are consistent within errors at $\sim$290~km~s$^{-1}$ in the UVES data and $\sim$270~km~s$^{-1}$ in the X-Shooter epoch. However, in the MagE spectrum (2021), we observe two distinct RV trends. Neutral atomic emission lines (e.g., Fe~\textsc{i}, Ti~\textsc{i}, Ni~\textsc{i}) are centered around $\sim$275~kms$^{-1}$, similar to the X-Shooter data, while absorption lines from Sc~\textsc{ii}, Si~\textsc{ii}, and the Paschen series indicate an RV of $317\pm7$~km~s$^{-1}$. This $\sim$40~km~s$^{-1}$ discrepancy suggests the presence of a binary system. Notably, the Paschen and Sc~\textsc{ii} lines are typically observed in stars ranging from late-B to late-G spectral types. These features could originate from the YHG if it became slightly hotter after 2016, in which case the RV shift would reflect orbital motion. Alternatively, these lines may trace the companion star, representing the first direct spectroscopic detection of the binary. In either scenario, the RV shifts support the presence of a binary companion.


\subsection*{The binary system}\label{sec:binary_system}

The system currently exhibits both yellow-warm and hot spectral features. Given that the $I$-band brightness remains almost constant across the transition, we infer that the same object continues to dominate the luminosity post-transition. The near-IR spectral features observed before the transition were consistent with a RSG \citep{Ohnaka2024}, whereas the current features correspond to a warmer object. Thus, we conclude that the RSG evolved into a YHG, while the He~\textsc{i} lines originate either from a hot companion \citep[e.g.,][]{Cowley1969} or from an accretion disk around that companion \cite[e.g.,][]{Beristain2001}. Assuming a mass ratio as low as 0.1, the companion could be a $\sim$2.8~\msun~main-sequence B star \citep{Pecaut2013}. However, producing the observed optical hot features requires the companion to be comparable in luminosity to the YHG, implying that the companion should be a late-O or early-B main-sequence star, yielding a mass ratio $\gtrsim 0.5$ \citep{Pecaut2013}. Observational surveys show that $\sim$20\% of RSGs in the LMC and SMC have hot companions \citep{Neugent2019, Neugent2020, Patrick2022}, supporting the feasibility of such a scenario. 

The semi-regular photometric variability observed during the 1990s, with a period of $\sim$850 days and modulated by changes in $A_V$, may trace the co-rotation of the dusty structure within the orbital plane. Variations in extinction could arise from azimuthal asymmetries within a toroidal dust configuration. A hot companion might locally destroy dust, producing brightness maxima at specific orbital phases when it passes between the primary and our line of sight. However, a Keplerian orbit with a period of 850~days around a 28~\msun, 1500~\rsun~RSG would place the companion within the stellar envelope \citep{Omni2025}. This issue could be alleviated by orbital eccentricity, allowing the companion to remain outside the envelope except during periastron passages. 

Alternatively, the 850-day signal may represent the half-period, implying a true orbital period of $\sim$1700 days. This would relax the orbital constraints and better accommodate the observed features. During mass transfer, material can be ejected through the Lagrange points, forming a spiral dust structure \citep{Shu1979}. In this context, photometric maxima could correspond to phases when the observer's line of sight passes between spiral arms, reducing the intervening extinction. Initially, \citet{Ohnaka2008} modeled the dust torus assuming a pole-on geometry, while more recent data suggest an edge-on orientation \citep{Ohnaka2024}. Notably, the observations by \citet{Ohnaka2008} were taken during a brightness maximum, implying that the "cavity" interpreted as a pole-on structure might instead be a low-extinction region, such as a gap between spiral arms.

Binary population models suggest that a 12~\msun~RSG in a system with an initial orbital period shorter than 2000 days will undergo mass transfer, regardless of the companion’s mass \citep{Ercolino2024}. For a 28~\msun~RSG, the higher luminosity and, consequently, larger radius imply that even longer orbital periods would be required to avoid mass transfer. Depending on the initial orbital period and the mass ratio, possible outcomes include a CE phase, atmospheric ejection, or a merger. Continued monitoring of radial velocities for both components is essential to constrain the system's orbital parameters. 

We attempted to fit the UVES spectrum, when the primary star was still a Red Supergiant, using \texttt{MARCS} models \cite{Gustafsson2008} for the primary and adding a blackbody contribution to simulate the secondary. The goal was to search for an excess around 5,000~\r{AA} that could not be explained by \texttt{MARCS} models alone. For this test, we explored a range of secondary temperatures (8,000 K~$<$~\teff~$<$~50,000 K) and flux contributions relative to the RSG at 5500~\r{AA} ($10^{-3} <$ flux ratio $<$ 10). However, this approach failed to improve the fit significantly, suggesting that either the companion is too faint, embedded in dust, eclipsed by the extended RSG envelope along the line of sight, or possibly even within the primary's envelope in a common-envelope evolution scenario.


\subsection*{The cause of the transition}\label{sec:transition}

Very luminous RSGs with $\dot{M}\sim10^{-4}$~\msun~yr$^{-1}$ \citep[e.g.,][]{Yang2023} naturally evolve to a YHG phase before ending their lives \citep[see the effect of Yang+23 winds in][]{Zapartas2024}. However, the timescale of such evolution is larger than the dramatic transition in WOH~G64. Therefore we proposed several scenarios to explain the transition. 

The high obscuration of WOH~G64 ($A_{\rm V}>6$~mag) \citep{Levesque2009} arises from a torus-shaped, dusty-gaseous envelope, with an estimated mass of 3-9~\msun~\citep{Ohnaka2008}. This extreme obscuration entails a high IR excess, which leads to a high mass-loss rate, $\dot{M} > 10^{-4}$~\msun~yr$^{-1}$, compared to the LMC RSG population  \citep{Antoniadis2024}. Binary interactions might explain the massive envelope enshrouding WOH~G64 if they are the dominant mechanism for mass ejection. Stable mass transfer through Roche-lobe overflow might account for the high $\dot{M}$ \citep[e.g.,][]{Ercolino2024}. Then, depending on the orbital parameters and the mass ratio, the system might enter a CE phase. The CE would explain the historically very cool \teff~of WOH~G64 compared to other RSGs \citep{vanLoon2001, Levesque2009, Levesque2017}. During the CE phase, the system may experience a slow, self-regulated phase that can last hundreds of years. In this regime, the energy from orbital decay is efficiently transported to the surface and radiated away \cite{Clayton2017}, resulting in large-amplitude, repeating pulsations. These pulsations can eject shells every few decades, producing a time-averaged mass-loss rate of \Mdot~$\sim 10^{-3}$ \msun~yr$^{-1}$.

A pioneer study proposed that eruptions up to $\dot{M}\sim10^{-2}$~\msun~yr$^{-1}$ can occur in supergiants between $10,000 >T_{\rm eff} > 3,100$~K when the local super-Eddington luminosity exceeds the binding energy of the overlaying envelope \cite{Cheng2024}. These eruptions can lead to major changes in the structure of the star and its observed properties. As mentioned in the main text, Var~A experienced a similar transition when its $\sim$45~yr eruption ended and it returned to its quiescent state. We compare the transitions of WOH~G64 with Var~A in the Hertzsprung–Russell diagram in Fig.~\ref{fig:fig_HRD}. We also added the LMC population of YHGs \citep{Kourniotis2022, Humphreys2023}, YSGs \citep{Neugent2012}, and RSGs \citep{Antoniadis2024}, and various extreme Galactic YHGs with significant transitions such as $\rm{\rho}$~Cas \citep{Lobel2003, vanGenderen2019}, HR 8752 \citep{Nieuwenhuijzen2012} and IRC+10420 \citep{Oudmaijer1996, Nieuwenhuijzen2000} for comparison. We included the \textsc{mist} evolutionary tracks with initial rotation $v= 0.4 v_{\rm rot}$ and [Fe/H]$=-0.25$~dex \citep{Dotter2016, Choi2016}. For WOH~G64, we assumed $\log(L/\lsun)=5.45 \pm 0.05$ \citep{Ohnaka2008}, a minimum $T_{\rm eff}=3,200$~K \citep{Levesque2009} and a new $T_{\rm eff} = 4,800 \pm 300$~K based on the near-IR.

We also considered an alternative scenario, where the transition is the brightening after a previous merger. V838 Monocerotis is a peculiar binary that underwent an immense stellar explosion in 2002, leaving behind an expanding cool supergiant, a B3~V companion, and a dense envelope of expanding molecular matter \citep{Bond2003, Exter2016, Liimets2023}. The most accepted explanation is a merger in a triple system causing a luminous red nova \citep{Kaminski2021}. Since then, the cool supergiant has been gradually brightening and increasing the temperature over 17~yr, changing from an L-type ($T_{\rm eff}\sim2,300$~K) \cite{Evans2003} to a late-M ($T_{\rm eff}\sim3,300$~K) \cite{Liimets2023}. WOH~G64 shares several characteristics with V838 Monocerotis, such as a very cool $T_{\rm eff}$~compared to other RSGs, a dense surrounding envelope, and a companion. However, the two systems also have key differences. The photometric variability of WOH~G64 during its cool RSG phase was more extreme; it exhibited strong nebular emission and had a disk-like circumstellar morphology, in contrast to the spherical shell observed in V838 Monocerotis.

A merger should have occurred in a previous triple system during the last century in WOH~G64, expelling the material and obscuring the system. In this scenario, the transition could be similar to the brightening and increase in $T_{\rm eff}$~reported in V838 Monocerotis. We examined the APASS Catalog ($B$-band) within the Digital Access to a Sky Century at Harvard \citep[DASCH;][]{Grindlay2009} for detections of WOH~G64 over the last century. If a merger occurred, the unobscured progenitors might have been detected in the $B$-band photometric plates. There are five detections with $16.0<B<17.2$~mag between 1928 and 1953 located at 9" from WOH~G64. However, they might be associated with two stars within a 15" radius from WOH~G64 with $B\sim17$~mag (NOMAD Catalog) \cite{Zacharias2004}. Thus, we lack definitive evidence to support the merger hypothesis beyond speculation. 

\section*{Data availability}

The photometric datasets are publicly available in the respective survey repositories, except for the OGLE data, which can be accessed at Figshare \cite{MunozSanchez2024_figshare}. The spectra obtained with ESO facilities are available through the ESO archive, while the MagE spectrum is provided at the same Figshare link.

\section*{Acknowledgments}

GMS, MK, SdW, KA, AZB, EC, and GM acknowledge funding support from the European Research Council (ERC) under the European Union’s Horizon 2020 research and innovation programme (``ASSESS", Grant agreement No. 772086). EZ acknowledges support from the Hellenic Foundation for Research and Innovation (H.F.R.I.) under the “3rd Call for H.F.R.I. Research Projects to support Post-Doctoral Researchers” (Project No: 7933). This work has been co-funded by the National Science Centre, Poland, grant No. 2022/45/B/ST9/00243. We acknowledge helpful discussions with Nadejda Blagorodnova, Willem-Jan de Wit, Andrea Ercolino, Ylva Götberg, Despina Hatzidimitriou, Michaela Kraus, Evgenia Koumpia, Michalis Kourniotis, Rene Oudmaijer, and Lee Patrick. The work of KB is supported by NOIRLab, which is managed by the Association of Universities for Research in Astronomy (AURA) under a cooperative agreement with the U.S. National Science Foundation. This paper includes data gathered with the 6.5m Magellan Telescopes located at Las Campanas Observatory, Chile.

\section*{Author's contributions}

All authors helped with the interpretation of the data and provided comments on the manuscript. GMS has led the analysis of the data and is the main author of the text. MK and SdW contributed significantly to the analysis of the data. KA built the spectral energy distribution after the transition. AZB supervised the project and contributed significantly to the text. EZ led the discussion on stellar evolution and the binary scenario. EC performed the telluric cleaning of the near-infrared spectroscopy. GM provided insight on the B[e] analysis. KB obtained the MagE spectrum; AU and IS provided the time-series photometry from OGLE.

\section*{Competing interests}
The authors declare no competing interests.

\section*{Tables}
\begin{table}[!htb]
\caption{Optical spectral classifications of WOH G64} \label{tab:spectral_class}
\smallskip 
\renewcommand{\arraystretch}{1.4}
\begin{tabular}{@{\extracolsep{\fill}}lccc}
\hline \hline
Reference             &UT Date       & Spectral Class.\\
\hline \\[-12pt]
Elias et al. 1986 \cite{Elias1986}    & --             & M7.5 \\          
van Loon et al. 2005 \cite{vanLoon2005}  &1995 Dec 8-11  & M7.5e \\       
This work             &2007 Oct 28    & M6  \\
Levesque et al. 2009 \cite{Levesque2009}  &2008 Dec 10     & M5 \\ 
This work             &2016 Jul 27    & B[e] star \\
This work             &2021 Dec 17    & B[e] star \\

  \hline 

\hline
\end{tabular}
\end{table}

\begin{table}
\small
\caption{Log of the spectroscopic observations presented in this work\label{tab:spec_obs}}
\smallskip 
\begin{tabular}{l c c c c c}
\hline\hline
Telescope & Instrument	& UT Date & Exp. time & $R$\\ 
	&	&   & \small{(s)} &   \\ 
\hline \\[-9pt]

UT2 VLT 8m & \textnormal{UVES}  & 2007 Oct 28 & $2\times1800 + 1\times300$ & 57000 \\ 
UT2 VLT 8m & \textnormal{X-Shooter}  & 2016 July 27 & 624$^{a}$, 980$^{b,c}$ & 6500$^a$, 4200$^{b,c}$\\ 
Magellan 6.5m &\textnormal{MagE}	& 2021 Dec 17 & $3\times180 + 1\times240$ & 4000\\ 
\hline
\end{tabular}
\footnotetext{\textbf{Notes}: $^a$Vis, $^b$UVB, $^c$NIR}
\end{table}

\noindent



\renewcommand{\figurename}{Extended Data Fig.}
\renewcommand{\tablename}{Extended Data Table}
\setcounter{figure}{0}
\setcounter{table}{0}

\section*{Extended data}\label{sec3}

\begin{figure}[H]
        \centering
   	\resizebox{1.0\hsize}{!}{\includegraphics{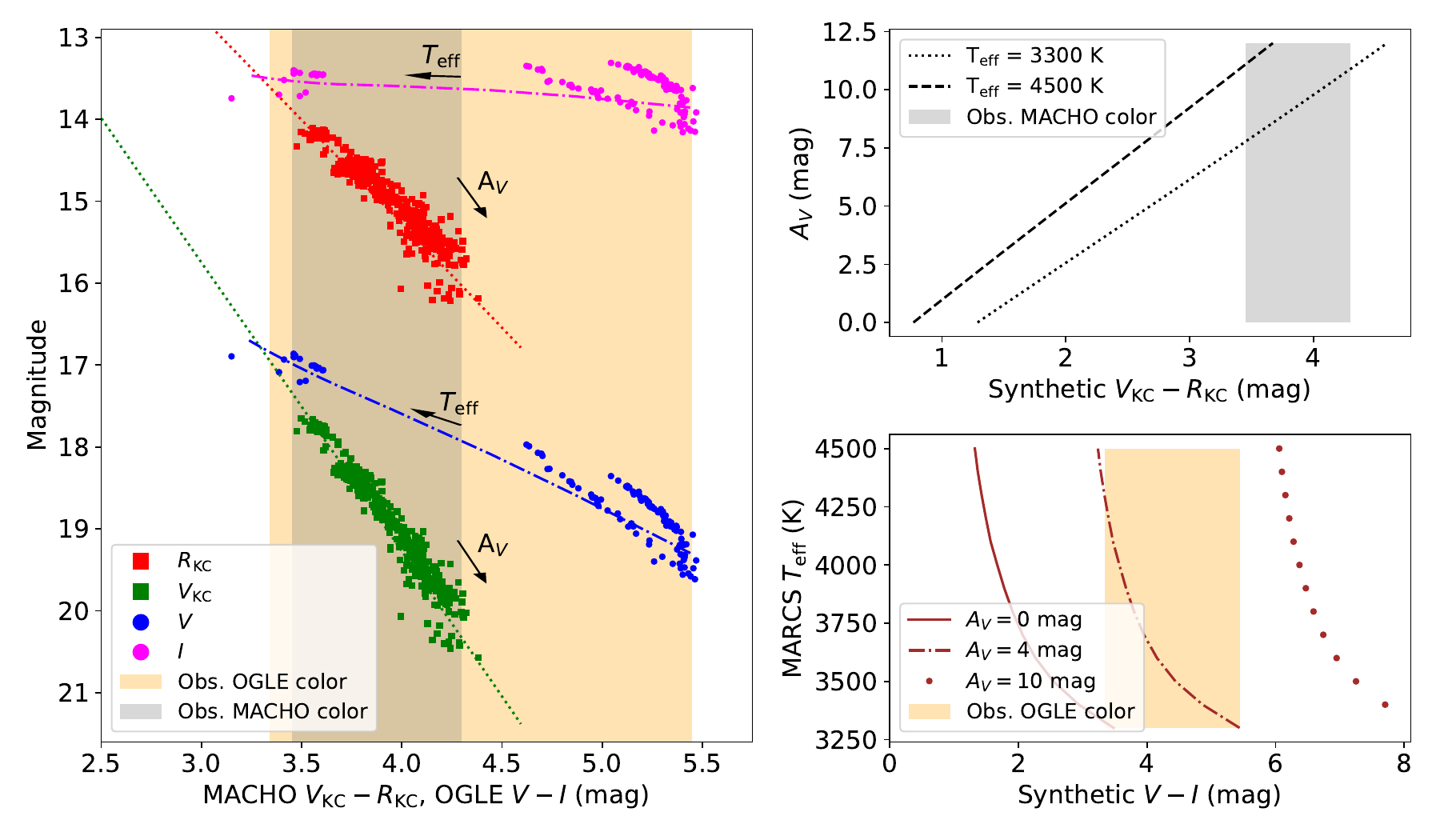}}
    \caption{\textbf{Comparison of observed color–magnitude variability with synthetic photometry.} Left: Photometric observations are shown with dots (OGLE, between January 2010 and October 2017) and squares (MACHO, between June 1993 and December 1999). The dotted line represents the MACHO synthetic colors for a model with $\teff = 3300$~K and variable $A_V$, while the dot-dashed line corresponds to OGLE synthetic colors for a model with variable $\teff$ and a fixed $A_V = 4$~mag. Black arrows indicate the direction of increasing $A_V$ and $T_{\rm eff}$ in the diagram. Shaded areas represent the full range of observed color variation. Top right: Values of $A_V$ required to reproduce the observed MACHO colors using synthetic photometry, as a function of $T_{\rm eff}$. Bottom right: Values of $T_{\rm eff}$ required to reproduce the observed OGLE colors using synthetic photometry, as a function of $A_V$.}
\label{fig:color_MARCS}
\end{figure}

\clearpage

\begin{figure}[H]
  \begin{center}
  \begin{subfigure}[b]{1.0\textwidth}

    \includegraphics[width=\textwidth]{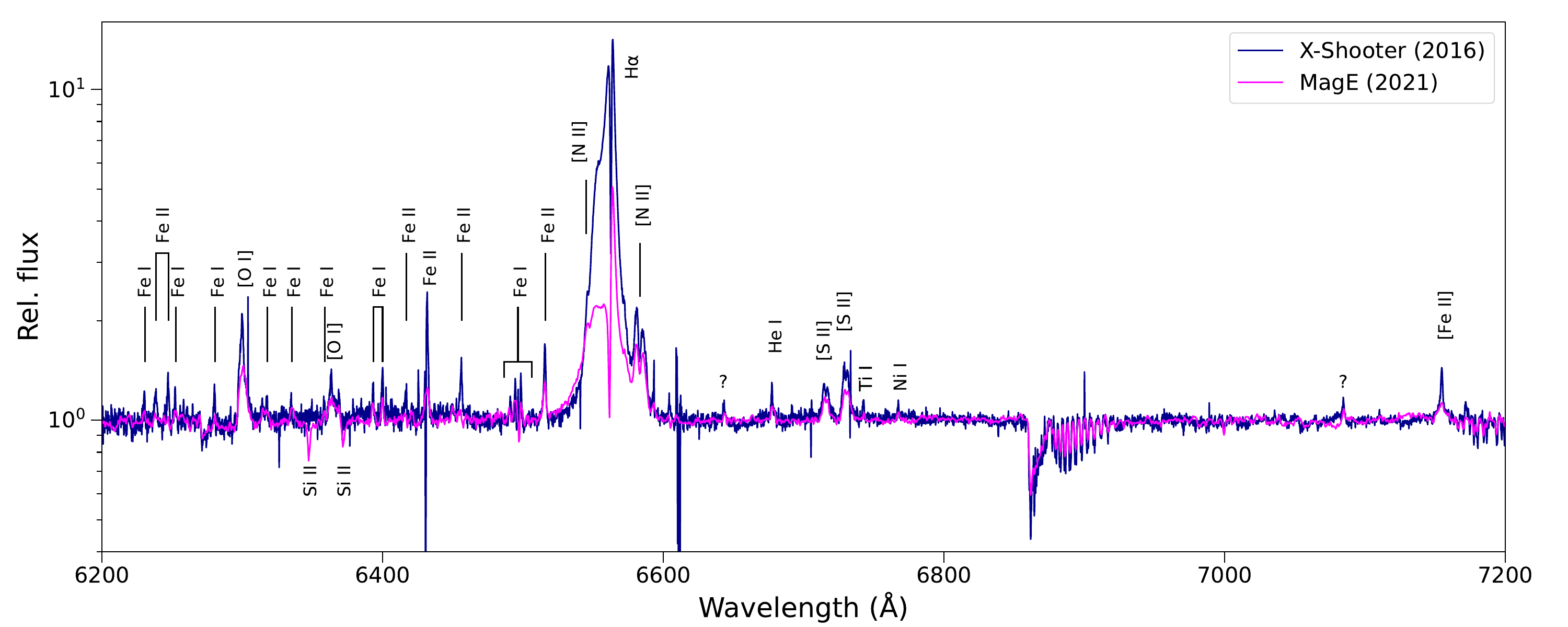}
    \label{fig:spec_pdf1}
  \end{subfigure}
  \begin{subfigure}[b]{1.0\textwidth}
  \centering
    \includegraphics[width=\textwidth]{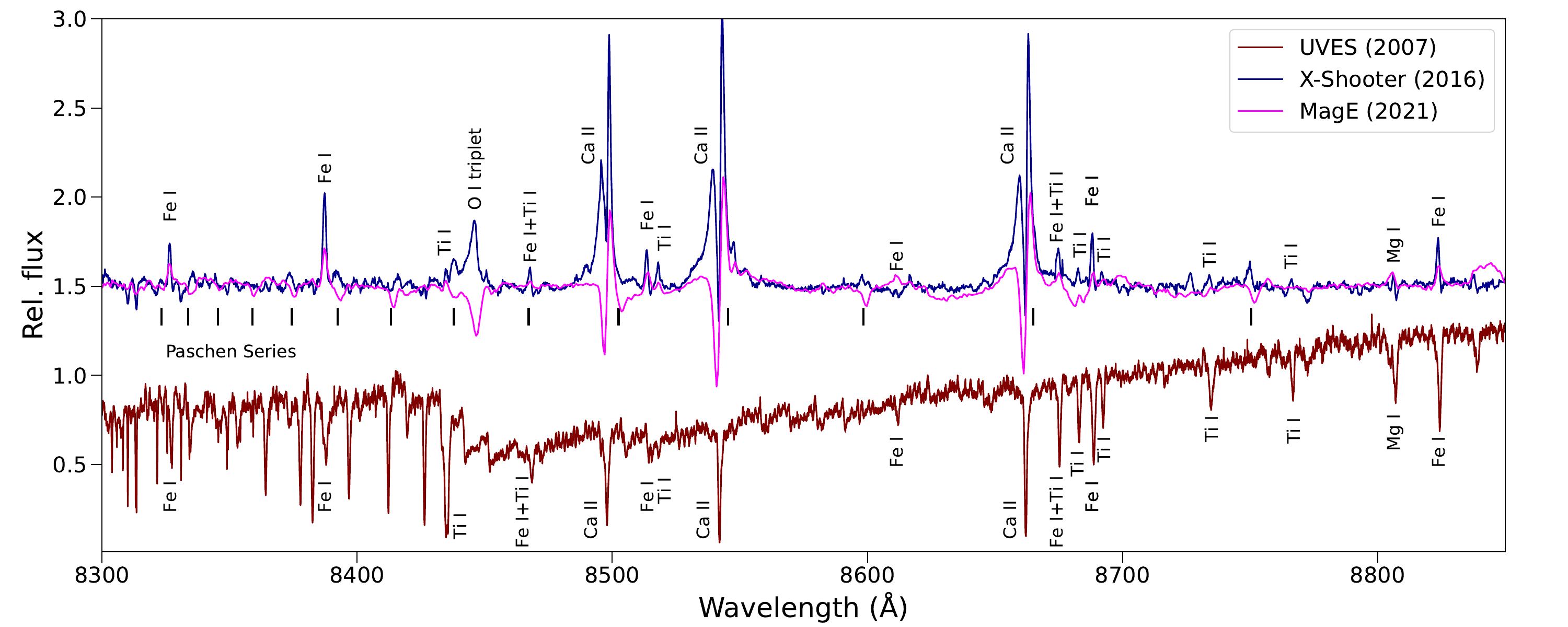}
    \label{fig:spec_pdf2}
  \end{subfigure}      
  \end{center}
    \caption{ \textbf{Comparison of the optical spectra of WOH~G64}. Top: Spectral region between 6200-7200~\r{A}. Bottom: Spectral region of the Ca~\textsc{ii} triplet region. The X-Shooter (2016) and MagE (2021) spectra are normalized, and the UVES spectrum is shown with an offset. The main spectral features are indicated for each epoch.}
    \label{fig:fig_spec_2016_2021_Ha_and_Ca_triplet}
\end{figure}

\clearpage

\begin{figure}[H]
    \centering
    \resizebox{0.95\hsize}{!}{\includegraphics{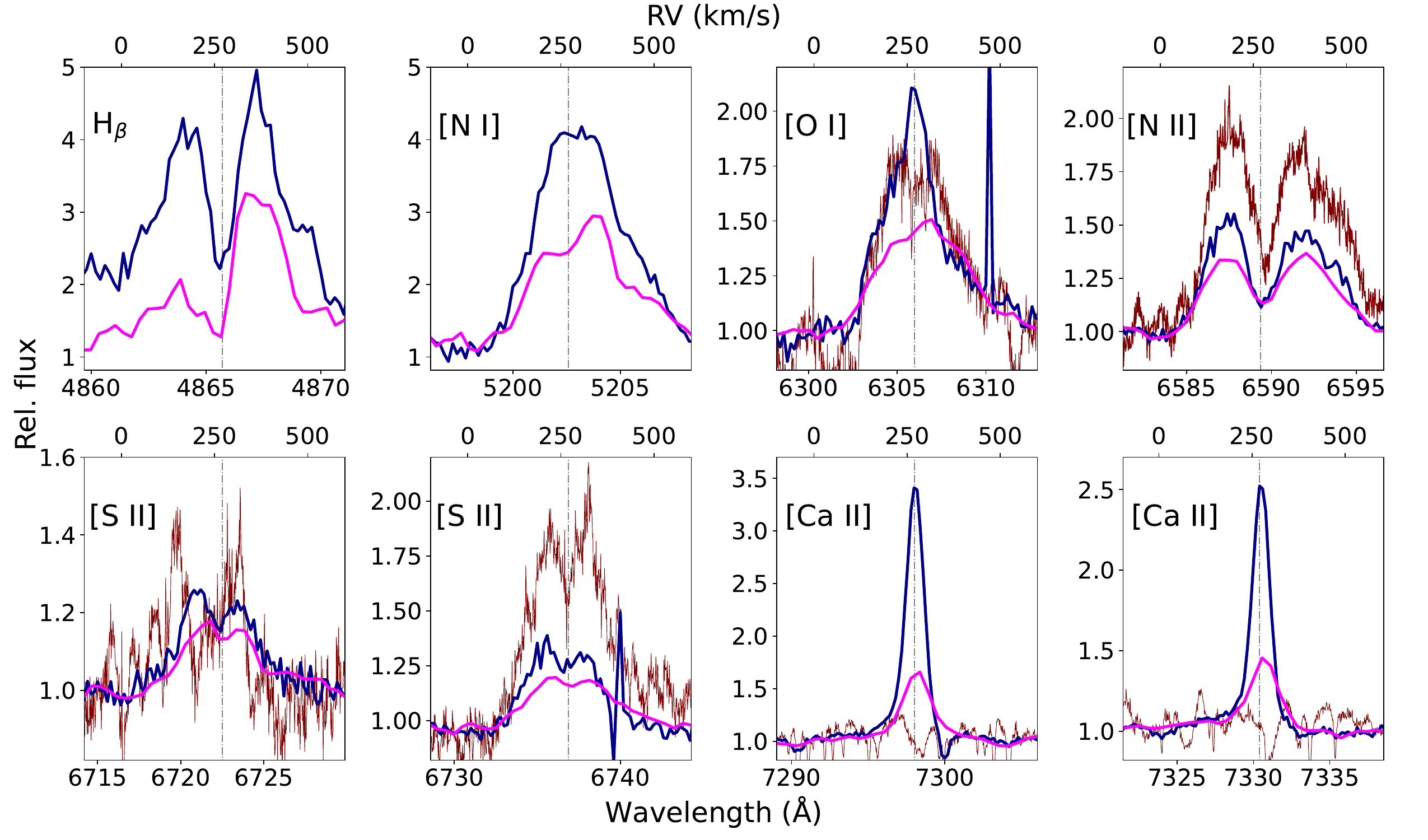}}
        \caption{\textbf{Comparison of the main forbidden lines found in WOH~G64 spectra}. The UVES spectrum is shown in brown, X-Shooter in blue, and MagE in magenta. The vertical gray dashed-dotted line indicates the center of the line assuming a RV of 270~km~s$^{-1}$.}
   \label{fig:fig_forbidden}
\end{figure}

\clearpage

\begin{figure}[H]
    \centering
    \resizebox{1.0\hsize}{!}{\includegraphics{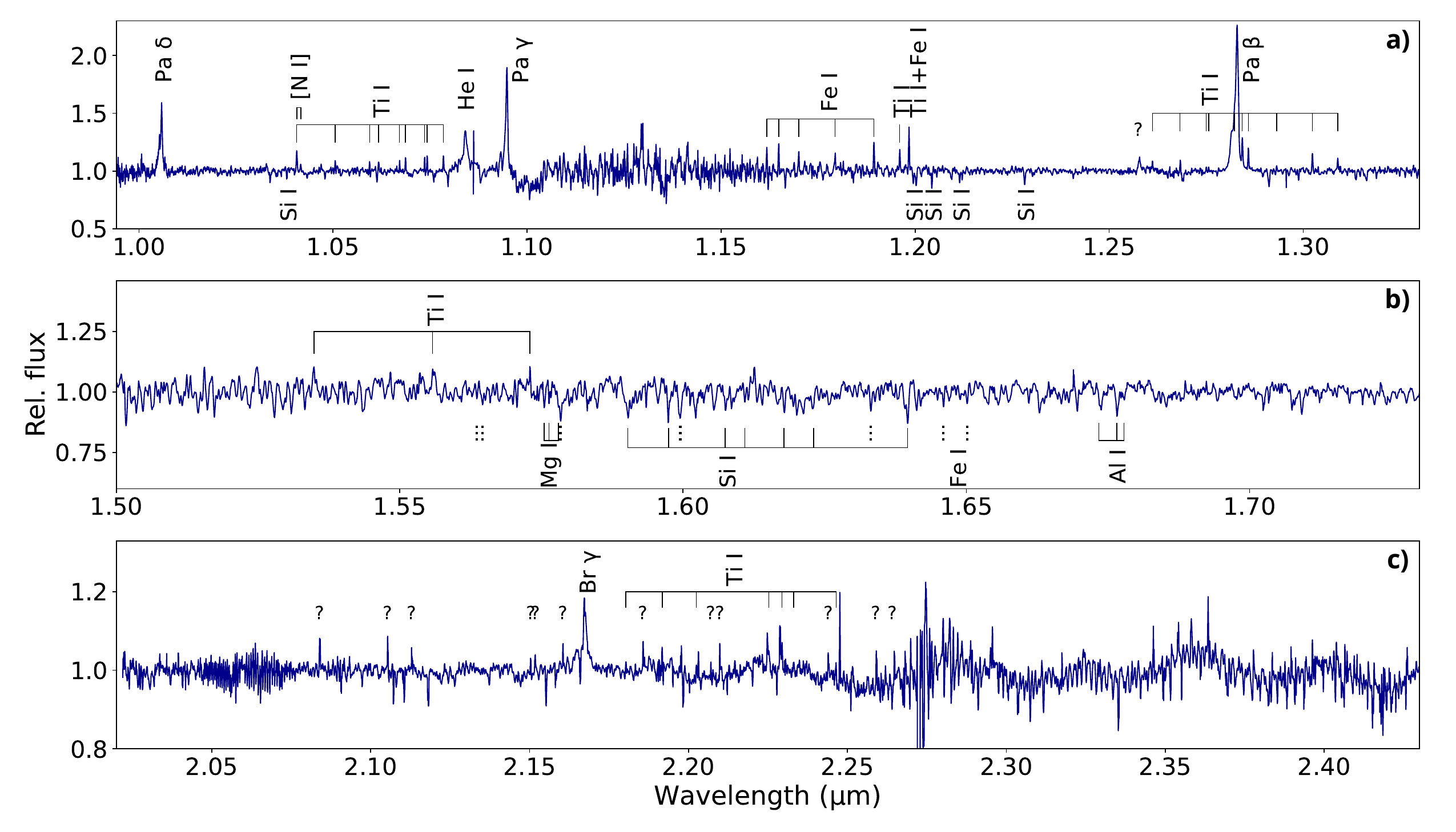}}
    \caption{\textbf{Near-infrared spectrum of WOH~G64 from X-Shooter (2016)}. The normalized spectrum of WOH~G64 is shown in $YJ$-band (top), $H$-band (middle), and $K$-band (bottom), indicating identified lines. Unidentified lines are marked with a '?'. }
    \label{fig:fig_spec_nearIR}
\end{figure}

\clearpage

\begin{figure}[H]
  \begin{center}
  \begin{subfigure}[b]{0.75\textwidth}

    \includegraphics[width=\textwidth]{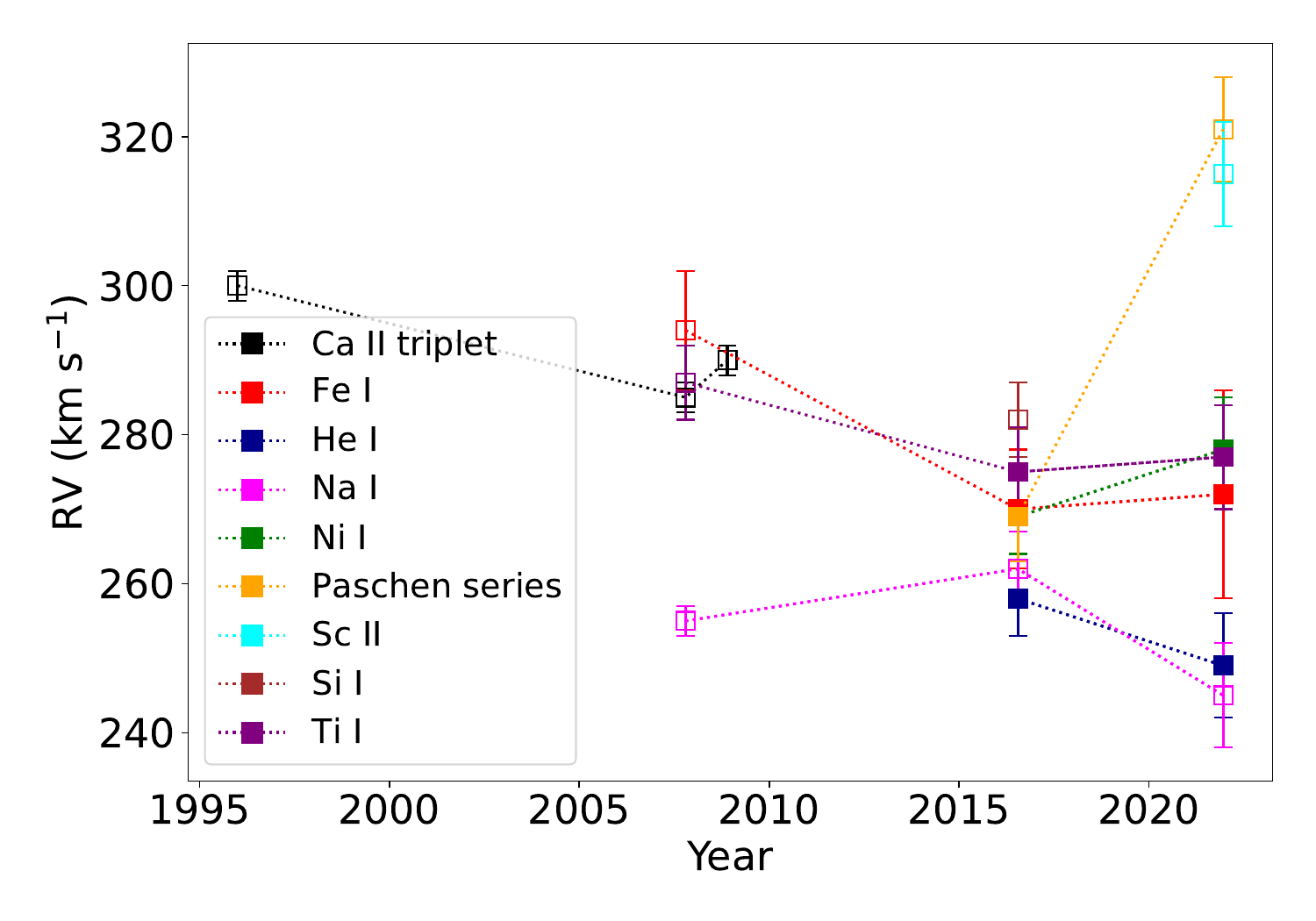}
    \label{fig:pdf1}
  \end{subfigure}
  \begin{subfigure}[b]{0.75\textwidth}
  \centering
    \includegraphics[width=\textwidth]{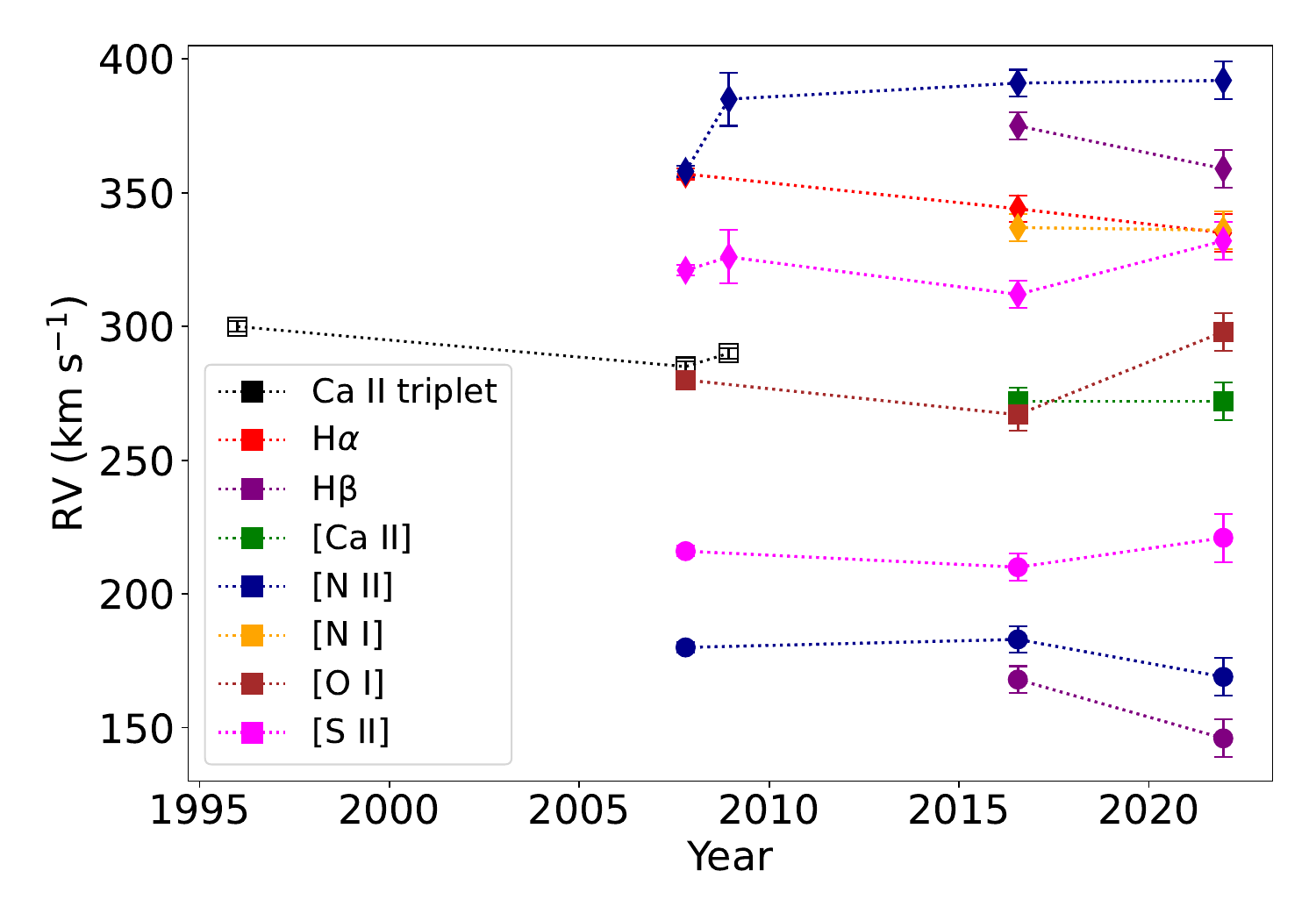}
    \label{fig:pdf2}
  \end{subfigure}      
  \end{center}
  \caption{\textbf{Radial velocity evolution of atomic lines with time}. Top: Median radial velocity of each ion. Filled squares indicate measurements from emission lines, and empty squares from absorption lines. Bottom: Median radial velocity of forbidden and double-peaked emission. The squares indicate single emission, the diamonds and circles are the red-shifted and blue-shifted peaks, respectively, in the double-peaked emission. Only the red-peaked emission from H$\alpha$ is shown. Ca~\textsc{ii} triplet measures are included as reference.}
  \label{fig:fig_RV}
\end{figure}

\clearpage

\begin{small}
\setlength\tabcolsep{3pt} 
\begin{longtable}{ccccc|ccccc}
\caption{Optical spectral lines of WOH G64 in the X-Shooter and MagE spectra} \label{tab:spectral_lines}

\smallskip 
\renewcommand{\arraystretch}{1.2}\\
\hline\hline\\[-7pt] 
$\lambda$ (\r{AA}) & Ion & UVES & X-Shooter & MagE & $\lambda$ (\r{AA}) & Ion & UVES & X-Shooter & MagE \\
\hline \\
4068.6 & [S~$\textsc{ii}$] & --- & Em & Em & 5641.0 & Sc~$\textsc{ii}$ & --- & --- & Abs\\
4340.46 & H${\rm\gamma}$ & --- & Em & Em & 5641.01 & Si~$\textsc{ii}$ & --- & --- & Abs\\
4861.32 & H${\rm\beta}$ & --- & double & double & 5657.91 & Sc~$\textsc{ii}$ & --- & --- & Abs\\
4921.93 & He~$\textsc{i}$ & --- & Em & Em & 5658.36 & Sc~$\textsc{ii}$ & --- & --- & Abs\\
5006.84 & [O~$\textsc{iii}$] & --- & x & Em & 5667.16 & Sc~$\textsc{ii}$ & --- & --- & Abs\\
5018.44 & Fe~$\textsc{ii}$ & --- & x & Abs & 5669.06 & Sc~$\textsc{ii}$ & --- & --- & Abs\\
5019.73 & Fe~$\textsc{i}$ & --- & Em & x & 5684.21 & Sc~$\textsc{ii}$ & --- & --- & Abs\\
5041.07 & Fe~$\textsc{i}$ & --- & Em & Em & 5889.95 & Na~$\textsc{i}$ & Abs & Abs & Abs\\
5051.63 & Fe~$\textsc{i}$ & --- & Em & x & 5895.92 & Na~$\textsc{i}$ & Abs & Abs & Abs\\
5079.9 & ? & --- & x & Abs & 5956.69 & Fe~$\textsc{i}$ & x & Em & Em\\
5079.74 & Fe~$\textsc{i}$ & --- & Em & x & 5991.37 & Fe~$\textsc{ii}$ & x & Em & Em\\
5083.34 & Fe~$\textsc{i}$ & --- & Em & x & 6065.48 & Fe~$\textsc{i}$ & x & Em & Em\\
5107.45 & Fe~$\textsc{i}$ & --- & Em & Em & 6084.1 & Fe~$\textsc{ii}$ & x & Em & Em\\
5123.72 & Fe~$\textsc{i}$ & --- & Em & x & 6108.12 & Ni~$\textsc{i}$ & x & Em & Em\\
5169.03 & Fe~$\textsc{ii}$ & --- & x & Abs & 6128.94 & ? & x & x & Abs\\
5183.49 & Fe~$\textsc{i}$ & --- & Em & x & 6126.22 & Ti~$\textsc{i}$ & x & x & Em\\
5183.81 & Ti~$\textsc{i}$ & --- & Em & x & 6137.69 & Fe~$\textsc{i}$ & x & Em & Em\\
5185.9 & Ti~$\textsc{ii}$ & --- & Em? & x & 6141.73 & Fe~$\textsc{i}$ & Abs & Em & Em\\
5197.9 & [N~$\textsc{i}$] & --- & Em & Em & 6141.71 & Ba~$\textsc{ii}$ & x & x & Abs\\
5216.27 & Fe~$\textsc{i}$ & --- & Em & x & 6191.56 & Fe~$\textsc{i}$ & x & Em & Em\\
5227.15 & Fe~$\textsc{i}$ & --- & Em & x & 6230.72 & Fe~$\textsc{i}$ & x & Em & Em\\
5234.62 & Fe~$\textsc{ii}$ & --- & Em & x & 6247.56 & Fe~$\textsc{ii}$ & x & Em & Em\\
5066.78 & ? & --- & Em & x & 6252.56 & Fe~$\textsc{i}$ & x & Em & Em\\
5269.54 & Fe~$\textsc{i}$ & --- & Em & x & 6256.36 & Ni~$\textsc{i}$ & x & Em & Em\\
5270.36 & Fe~$\textsc{i}$ & --- & Em & x & 6261.1 & Ti~$\textsc{i}$ & Abs & x & x\\
5273.37 & Fe~$\textsc{i}$ & --- & Em & x & 6280.62 & Fe~$\textsc{i}$ & x & Em & Em\\
5276.0 & Fe~$\textsc{ii}$ & --- & Em & x & 6300.3 & [O~$\textsc{i}$] & Em & Em & Em\\
5283.62 & Fe~$\textsc{i}$ & --- & Em & x & 6314.66 & Ni~$\textsc{i}$ & x & Em & Em\\
5316.61 & Fe~$\textsc{ii}$ & --- & Em & Abs? & 6318.02 & Fe~$\textsc{i}$ & x & Em & x\\
5328.04 & Fe~$\textsc{i}$ & --- & Em & x & 6335.33 & Fe~$\textsc{i}$ & x & Em & Em\\
5341.02 & Fe~$\textsc{i}$ & --- & Em & x & 6347.11 & Si~$\textsc{ii}$ & x & x & Abs\\
5367.65 & ? & --- & Em & x & 6358.7 & Fe~$\textsc{i}$ & x & Em & Em\\
5371.49 & Fe~$\textsc{i}$ & --- & Em & x & 6363.78 & [O~$\textsc{i}$] & x & Em & Em\\
5381.02 & Ti~$\textsc{ii}$ & --- & Em? & Abs? & 6369.45 & Fe~$\textsc{ii}$ & x & Em & x\\
5366.69 & ? & --- & Em & x & 6371.37 & Si~$\textsc{ii}$ & x & x & Abs\\
5397.13 & Fe~$\textsc{i}$ & --- & Em & x & 6393.6 & Fe~$\textsc{i}$ & x & Em & Em\\
5405.77 & Fe~$\textsc{i}$ & --- & Em & Em & 6400.32 & Fe~$\textsc{i}$ & Abs & Em & Em\\
5418.77 & Ti~$\textsc{ii}$ & --- & Em & x & 6416.92 & Fe~$\textsc{ii}$ & x & Em & x\\
5429.7 & Fe~$\textsc{i}$ & --- & Em & x & 6421.35 & Fe~$\textsc{i}$ & Abs & x & Em\\
5434.52 & Fe~$\textsc{i}$ & --- & Em & x & 6432.67 & Fe~$\textsc{ii}$ & x & P-Cygni & Em\\
5446.92 & Fe~$\textsc{i}$ & --- & Em & x & 6456.38 & Fe~$\textsc{ii}$ & x & Em & x\\
5455.61 & Fe~$\textsc{i}$ & --- & Em & Em & 6491.57 & Ti~$\textsc{ii}$ & x & Em & x\\
5526.78 & Sc~$\textsc{ii}$ & --- & --- & Abs & 6494.98 & Fe~$\textsc{i}$ & x & Em & Em\\
5577.37 & ? & --- & Em & x & 6496.9 & Ba~$\textsc{ii}$ & Abs & x & x\\
\hline\\
6498.94 & Fe~$\textsc{i}$ & x & Em & Em & 8345.55 & Pa20 & x & x & Abs\\
6516.08 & Fe~$\textsc{ii}$ & x & Em & Em & 8359.0 & Pa19 & x & x & Abs\\
6548.05 & [N~$\textsc{ii}$] & x & Em & Em & 8387.77 & Fe~$\textsc{i}$ & Abs & Em & Em\\
6562.8 & H$\alpha$ & double & double & P-Cygni & 8374.48 & Pa18 & x & x & Abs\\
6572.78 & Ca~$\textsc{i}$ & Abs & x & x & 8392.4 & Pa17 & x & x & Abs\\
6583.45 & [N~$\textsc{ii}$] & double & double & double & 8413.32 & Pa16 & x & x & Abs\\
6592.91 & Fe~$\textsc{i}$ & x & Em & Em & 8434.95 & Ti~$\textsc{i}$ & x & Em & Em\\
6649.8 & ? & Abs & Em & Em & 8437.96 & Pa15 & x & x & blended\\
6663.44 & Fe~$\textsc{i}$ & x & x & Em & 8446.36 & O~$\textsc{i}$ & x & Em & Abs\\
6678.15 & He~$\textsc{i}$ & x & Em & Em & 8467.25 & Pa14 & x & x & blended\\
6716.44 & [S~$\textsc{ii}$] & x & double & double & 8498.02 & Ca~$\textsc{ii}$ & Abs & double & P-Cygni\\
6730.81 & [S~$\textsc{ii}$] & Em & double & double & 8502.48 & Pa13 & x & blended & blended\\
6743.12 & Ti~$\textsc{i}$ & x & Em & Em & 8514.07 & Fe~$\textsc{i}$ & Abs & Em & Em\\
6767.77 & Ni~$\textsc{i}$ & x & Em & Em & 8518.19 & Ti~$\textsc{i}$ & Abs & Em & Em\\
6945.2 & Fe~$\textsc{i}$ & x & x & Em & 8542.09 & Ca~$\textsc{ii}$ & Abs & double & P-Cygni\\
6999.88 & Fe~$\textsc{i}$ & x & Em & Em & 8545.38 & Pa12 & x & blended & blended\\
7091.5 & ? & x & Em & Em & 8582.91 & Cr~$\textsc{i}$ & x & Abs & Em\\
7110.9 & Ni~$\textsc{i}$ & x & Em & x & 8598.39 & Pa11 & x & Em & Abs\\
7155.17 & [Fe~$\textsc{ii}$] & x & Em & Em & 8611.8 & Fe~$\textsc{i}$ & Abs & x & Em\\
7291.47 & [Ca~$\textsc{ii}$] & x & Em & Em & 8616.95 & [Fe~$\textsc{ii}$] & x & Em & x\\
7323.89 & [Ca~$\textsc{ii}$] & x & Em & Em & 8662.14 & Ca~$\textsc{ii}$ & Abs & double & P-Cygni\\
7462.41 & Fe~$\textsc{ii}$ & x & Em & Em & 8665.02 & Pa10 & x & blended & blended\\
7664.91 & K~$\textsc{i}$ & --- & Abs & Abs & 8674.75 & Fe~$\textsc{i}$ & Abs & Em & Em\\
7698.97 & K~$\textsc{i}$ & Abs & Abs & Abs & 8675.37 & Ti~$\textsc{i}$ & Abs & Em & Em\\
7711.72 & Fe~$\textsc{ii}$ & x & Em & Em & 8688.5 & Fe~$\textsc{i}$ & Abs & Em & Em\\
7714.32 & Ni~$\textsc{i}$ & x & Em & Em & 8692.33 & Ti~$\textsc{i}$ & Abs & Em & Em\\
7748.27 & Fe~$\textsc{i}$ & x & x & Em & 8735.04 & ? & x & Em & x\\
7774.17 & O~$\textsc{i}$ & x & x & Abs & 8734.71 & Ti~$\textsc{i}$ & Abs & Em & x\\
7788.94 & Ni~$\textsc{i}$ & x & Em & Em & 8750.47 & Pa9 & x & Em & Abs\\
7912.87 & Fe~$\textsc{i}$ & x & Em & Em & 8757.19 & Fe~$\textsc{i}$ & Abs & x & Em\\
7978.82 & Ti~$\textsc{i}$ & x & Em & Em & 8764.27 & Fe~$\textsc{ii}$ & x & Abs & x\\
8047.62 & Fe~$\textsc{i}$ & Abs & Em & Em & 8766.68 & Ti~$\textsc{i}$ & Abs & Em & x\\
8068.24 & Ti~$\textsc{i}$ & x & Em & Em & 8824.22 & Fe~$\textsc{i}$ & Abs & Em & Em\\
8075.15 & Fe~$\textsc{i}$ & Abs & Em & Em & 8862.78 & Pa8 & x & Em & Abs\\
8183.26 & Na~$\textsc{i}$ & Abs & x & x & 8921.76 & ? & x & x & Abs\\
8194.82 & Na~$\textsc{i}$ & Abs & x & x & 9014.91 & Pa7 & x & Em & Abs\\
8327.06 & Fe~$\textsc{i}$ & Abs & Em & Em & 9229.01 & Pa6 & x & Em & Abs\\
8327.06 & Fe~$\textsc{ii}$ & x & Em & x &  &  &  &  & \\
\hline \\


\end{longtable}
\textbf{Notes:} Emission lines are indicated with "Em", absorption lines with "Abs", P Cygni profiles with "P-Cygni", and double with "double". When the profile was irregular due to blending with other lines, we used the label "blended". The "x" indicates the absence of emission or absorption, while "---" means that the spectrum did not cover the line. We also report the observed wavelength of unidentified lines ("?").
\end{small}

\begin{small}
\begin{table}[!htb]

\caption{Near-IR spectral lines of WOH G64 in the X-Shooter spectrum} \label{tab:spectral_lines_nearIR}
\smallskip 
\renewcommand{\arraystretch}{1.25}
\begin{tabular}{ccc|ccc}
\hline\hline\\[-7pt] 
Wavelength ($\mu$m) & Ion & X-Shooter & Wavelength ($\mu$m) & Ion & X-Shooter \\
\hline
0.9546 & Pa$\epsilon$ & Em & 1.56217 & Fe~$\textsc{i}$ & Abs\\
0.97182 & [Fe~$\textsc{ii}$] & Em & 1.56319 & Fe~$\textsc{i}$ & Abs\\
1.00494 & Pa$\delta$ & Em & 1.57156 & Ti~$\textsc{i}$ & Abs\\
1.03713 & Si~$\textsc{i}$ & Abs & 1.57407 & Mg~$\textsc{i}$ & Abs\\
1.03968 & Ti~$\textsc{i}$ & Em & 1.57490 & Mg~$\textsc{i}$ & Abs\\
1.03979 & [N~$\textsc{i}$] & Em & 1.57658 & Mg~$\textsc{i}$ & Abs\\
1.04074 & [N~$\textsc{i}$] & Em & 1.57694 & Fe~$\textsc{i}$ & Abs\\
1.04961 & Ti~$\textsc{i}$ & Em & 1.58884 & Si~$\textsc{i}$ & Abs\\
1.05846 & Ti~$\textsc{i}$ & Em & 1.59600 & Si~$\textsc{i}$ & Abs\\
1.06034 & Si~$\textsc{i}$ & Abs & 1.59807 & Fe~$\textsc{i}$ & Abs\\
1.06077 & Ti~$\textsc{i}$ & Em & 1.60600 & Si~$\textsc{i}$ & Abs\\
1.06616 & Ti~$\textsc{i}$ & Em & 1.60948 & Si~$\textsc{i}$ & Abs\\
1.06770 & Ti~$\textsc{i}$ & Em & 1.61637 & Si~$\textsc{i}$ & Abs\\
1.07263 & Ti~$\textsc{i}$ & Em & 1.62157 & Si~$\textsc{i}$ & Abs\\
1.07329 & Ti~$\textsc{i}$ & Em & 1.63163 & Fe~$\textsc{i}$ & Abs\\
1.07747 & Ti~$\textsc{i}$ & Em & 1.63816 & Si~$\textsc{i}$ & Abs\\
1.08303 & He~$\textsc{i}$ & Em & 1.64448 & Fe~$\textsc{i}$ & Abs\\
1.09381 & Pa$\gamma$ & Em & 1.64867 & Fe~$\textsc{i}$ & Abs\\
1.16076 & Fe~$\textsc{i}$ & Em & 1.67190 & Al~$\textsc{i}$ & Abs\\
1.16383 & Fe~$\textsc{i}$ & Em & 1.67504 & Al~$\textsc{i}$ & Abs\\
1.16900 & Fe~$\textsc{i}$ & Em & 1.67505 & Al~$\textsc{i}$ & Abs\\
1.17833 & Fe~$\textsc{i}$ & Em & 1.67633 & Al~$\textsc{i}$ & Abs\\
1.18828 & Fe~$\textsc{i}$ & Em & 2.17829 & Ti~$\textsc{i}$ & Em\\
1.19495 & Ti~$\textsc{i}$ & Em & 2.18974 & Ti~$\textsc{i}$ & Em\\
1.19842 & Si~$\textsc{i}$ & Abs & 2.20045 & Ti~$\textsc{i}$ & Em\\
1.19916 & Si~$\textsc{i}$ & Abs & 2.22328 & Ti~$\textsc{i}$ & Em\\
1.21035 & Si~$\textsc{i}$ & Abs & 2.22740 & Ti~$\textsc{i}$ & Em\\
1.22707 & Si~$\textsc{i}$ & Abs & 2.23106 & Ti~$\textsc{i}$ & Em\\
1.25668 & [Fe~$\textsc{ii}$] & Em & 2.24439 & Ti~$\textsc{i}$ & Em\\
1.26003 & Ti~$\textsc{i}$ & Em & 2.08396 & ? & Em\\
1.26711 & Ti~$\textsc{i}$ & Em & 2.10527 & ? & Em\\
1.27384 & Ti~$\textsc{i}$ & Em & 2.11290 & ? & Em\\
1.27450 & Ti~$\textsc{i}$ & Em & 2.15031 & ? & Em\\
1.28037 & Fe~$\textsc{i}$ & Em & 2.15165 & ? & Em\\
1.28181 & Pa$\beta$ & Em & 2.16040 & ? & Em\\
1.28314 & Ti~$\textsc{i}$ & Em & 2.18562 & ? & Em\\
1.28470 & Ti~$\textsc{i}$ & Em & 2.20690 & ? & Em\\
1.29200 & Ti~$\textsc{i}$ & Em & 2.20973 & ? & Em\\
1.30119 & Ti~$\textsc{i}$ & Em & 2.24386 & ? & Em\\
1.30773 & Ti~$\textsc{i}$ & Em & 2.25897 & ? & Em\\
1.53348 & Ti~$\textsc{i}$ & Abs & 2.26462 & ? & Em\\
1.53348 & Ti~$\textsc{i}$ & Abs &  &  & \\
\hline 
\end{tabular}
\footnotetext{\textbf{Notes}: The notation is identical to that in Supplementary Material Table~\ref{tab:spectral_lines}.}
\end{table}

\end{small}

\bigskip





\clearpage
\bibliography{sn-bibliography}








\end{document}